\pdfoutput=1
\documentclass[twocolumn,english,aps,superscriptaddress,prb]{revtex4-1}

\usepackage{babel}
\usepackage{amsmath}
\usepackage{amssymb}
\usepackage{graphicx}

\begin{document}

\title{Floating, critical and dimerized phases in a frustrated spin-3/2 chain}

\author{Natalia Chepiga}
\affiliation{Institute for Theoretical Physics, University of Amsterdam, Science Park 904 Postbus 94485, 1090 GL Amsterdam, The Netherlands}
\author{Ian Affleck}
\affiliation{Department of Physics and Astronomy, University of British Columbia, Vancouver, BC, Canada V6T 1Z1}
\author{Fr\'ed\'eric Mila}
\affiliation{Institute of Physics, Ecole Polytechnique F\'ed\'erale de Lausanne (EPFL), CH-1015 Lausanne, Switzerland}

\date{\today}
\begin{abstract} 
We study spontaneous dimerization and emergent criticality in a spin-3/2 chain with antiferromagnetic nearest-neighbor $J_1$, next-nearest-neighbor $J_2$ and three-site $J_3$ interactions. In the absence of three-site interaction $J_3$, we provide evidence that the model undergoes a remarkable sequence of three phase transitions as a function of $J_2/J_1$, going successively through a critical commesurate phase, a partially dimerized gapped phase, a critical floating phase with quasi-long-range incommensurate order, to end up in a fully dimerized phase at very large $J_2/J_1$. In the field theory language, this implies that the coupling constant of the marginal operator responsible for dimerization changes sign three times. For large enough $J_3$, the fully dimerized phase is stabilized for all $J_2$, and the phase transitions between the critical phases and this phase are both Wess-Zumino-Witten (WZW) SU(2)$_3$ along part of the boundary and turn first order at some point due to the presence of a  marginal operator in the WZW SU(2)$_3$ model. By contrast, the transition between the two dimerized phase is always first order, and the phase transitions between the partially dimerized phase and the critical phases are Kosterlitz-Thouless. Finally, we discuss the intriguing spin-1/2 edge states that emerge in the partially dimerized phase for even chains. Unlike their counterparts in the spin-1 chain, they are not confined and disappear upon increasing $J_2$ in favour of a reorganization of the dimerization pattern.
\end{abstract}

\maketitle


\section{Introduction}

Antiferromagnetic  Heisenberg  spin  chains  have  attracted a lot of attention  over  the  years.   
Competing interactions induce frustration and are known to lead to new phases and quantum phase transitions.
For example, the $J_1-J_2$ spin-1/2 chain undergoes spontaneous dimerization \cite{MajumdarGhosh} when the ratio of the next-nearest neighbor interaction to the nearest neighbor one $J_2/J_1>0.2411$\cite{okamoto}. 
In the spin-1 chain, spontaneous dimerization is known to be induced by a negative biquadratic interaction $J_\mathrm{biq}/J_1<-1$. Recently, it has been shown that the three-site interaction $J_3\left[({\bf S}_{i-1}\cdot{\bf S}_{i})({\bf S}_{i}\cdot{\bf S}_{i+1})+\mathrm{h.c} \right]$ induces a fully dimerized state in spin-S chains and reduces to the $J_1-J_2$ model for spin-1/2\cite{michaud1}. Each of these two terms leads, when combined with nearest and next-nearest-neighbor interaction, to a rich phase diagram for the spin-1 chain. In particular, the quantum phase  transition between the next-nearest-neighbor (NNN-) Haldane phase that appears at large $J_2$ coupling \cite{kolezhuk_prl,kolezhuk_prb,kolezhuk_connectivity} and the dimerized phase has been shown to be in the Ising universality class with a singlet-triplet gap that remains open\cite{j1j2j3_short,j1j2jb_comment,j1j2j3_long}. Besides, the Wess-Zumino-Witten (WZW) SU(2)$_2$ critical line between Haldane and dimerized phases turns into a first order transition due to the presence of a marginal operator in the underlying critical theory. 

The Heisenberg spin-1 chain has a bulk gap\cite{Haldane}, but spin-1/2 edge states. In finite systems, the coupling between these two edge spins decay exponentially fast with the length of the chain. This causes quasi-degenerate low-lying in-gap states, a singlet and the so-called Kennedy triplet\cite{Kennedy}. By contrast, half-integer-spin chains with isotropic nearest-neighbor Heisenberg interaction ($J_2=J_3=0$) are known to be gappless with algebraically decaying spin-spin correlations. Edge spins do not appear in the critical spin-1/2 chain, although they emerge for higher spins. 
Edge states in critical systems are fundamentally different from those in the gapped phase: in the critical spin-3/2 chain the spin-1/2 edge states are logarithmically delocalized over the entire chain, moreover the energy splitting between the singlet and triplet low-lying states scales with the length of the chain L as $\Delta_{\mathrm edge}\propto 1/(L \ln (BL))$, where $B$ is a non-universal constant\cite{igloi}.

A previous investigation\cite{roth}  of the spin-3/2  $J_1-J_2$ chain has shown that the system undergoes spontaneous dimerization when $J_2/J_1>0.29$. The transition between the critical and the gapped dimerized phases is expected to be in the WZW SU(2)$_1$ universality class in analogy with the spin-1/2 $J_1-J_2$ model. In this gapped phase, emergent spin-1/2 edge states are localized at the open ends of a chain. The appearance of these edge states in the gapped dimerized phase suggests that it is rather partially dimerized with alternating single and double valence bond singlets (VBS) between nearest-neighbor sites. The edge states disappear around $J_2/J_1\approx0.48$\cite{roth}. Since the correlation length remains finite, it has been proposed that the system undergoes a first order phase transition at that ratio by analogy with the spin-1 $J_1-J_2$ model\cite{kolezhuk_prl}.  

In the spin-3/2 $J_1-J_3$ model the transition from the critical WZW SU(2)$_1$ phase\cite{affleck_haldane} to a spontaneously dimerized one occurs at $J_3/J_1\approx 0.063$\cite{michaud2}. This phase transition is continuous and belongs to the SU(2)$_{k=3}$ WZW universality class characterized by the central charge $c=2$.
This is in agreement with a recent prediction on symmetry protection of critical phases\cite{FuruyaOshikawa} that a renormalization-group flow from WZW SU(2)$_{k_1}$ to SU(2)$_{k_2}$ theory is possible only if the parity from $k_1$ to $k_2$ is preserved.

The dimerized phase induced by the three-site interaction corresponds to a fully dimerized phase with three VBS singlets on every other nearest-neighbor bond. 
 To see this we refer to a special point in $J_1-J_3$ model where the ground state is known exactly. Michaud et al.\cite{michaud1,michaud2} have shown that at $J_3/J_1=1/(4S(S+1)-2)$ the ground-state is an exactly dimerized state for all spin-S chains.
Further investigations have shown that this exact result can be extended to the case where a next-nearest neighbor exchange $J_2$  is included\cite{wang}. The two fully dimerized states are eigenstates along the line 
\begin{equation}
\frac{J_3}{J_1-2J_2}=\frac{1}{4S(S+1)-2},
\label{nnn_dimerization}
\end{equation}
and they are ground states for $J_2$ not too large. For spin-3/2, eq.\ref{nnn_dimerization} implies $J_3/(J_1-2J_2)=1/13$.

In the present paper we study the combined effect of next-nearest-neighbor and three-site interactions in the spin-3/2 chain. The model is defined by the $J_1-J_2-J_3$ Hamiltonian:
\begin{multline}
  H=J_1\sum_i {\bf S}_i\cdot{\bf S}_{i+1}+J_2\sum_i {\bf S}_{i-1}\cdot{\bf S}_{i+1}\\
  +J_3\sum_i\left[({\bf S}_{i-1}\cdot {\bf S}_i)({\bf S}_i\cdot {\bf S}_{i+1})+{\mathrm h.c.}\right].
  \label{eq:j1j2j3s}
\end{multline}
In the following, we focus on $J_1,J_2,J_3>0$, and without loss of generality 
we fix $J_1=1$. The numerical simulations have been performed with a state-of-the art density matrix renormalization group (DMRG) algorithm\cite{dmrg1,dmrg2,dmrg3,dmrg4}. Throughout the paper we consider chains of even length with open boundary conditions.

The paper is organized as follows. In Sec.~\ref{sec:phase_diagram} we provide an overview of the phase diagram and discuss its main features. In Sec.~\ref{sec:CCtoFD} we focus on small values of $J_2$ and study the nature of the phase transition between the commensurate-critical phase and the fully dimerized one. In Sec.~\ref{sec:KTtransition_s15} we provide numerical evidence for a Kosterlitz-Thouless transition between the commensurate-critical and partially dimerized phases. In the following Sec.~\ref{sec:dimer_to_dimer} we provide numerical evidence in favor of a first order transition between the two dimerized phases. We study the critical incommensurate phase that emerges at large values of $J_2$ and discuss the nature of the phase transitions between this floating phase and the two dimerized phases in Sec.~\ref {sec:floating}. In Sec.~\ref{sec:edge states} we discuss the behavior of the edge states in the partially dimerized phase. Sec.~\ref{sec:conclusion} contains our final discussion and conclusions.


\section{Phase diagram}
\label{sec:phase_diagram}

\begin{figure}[h!]
\centering 
\includegraphics[width=0.5\textwidth]{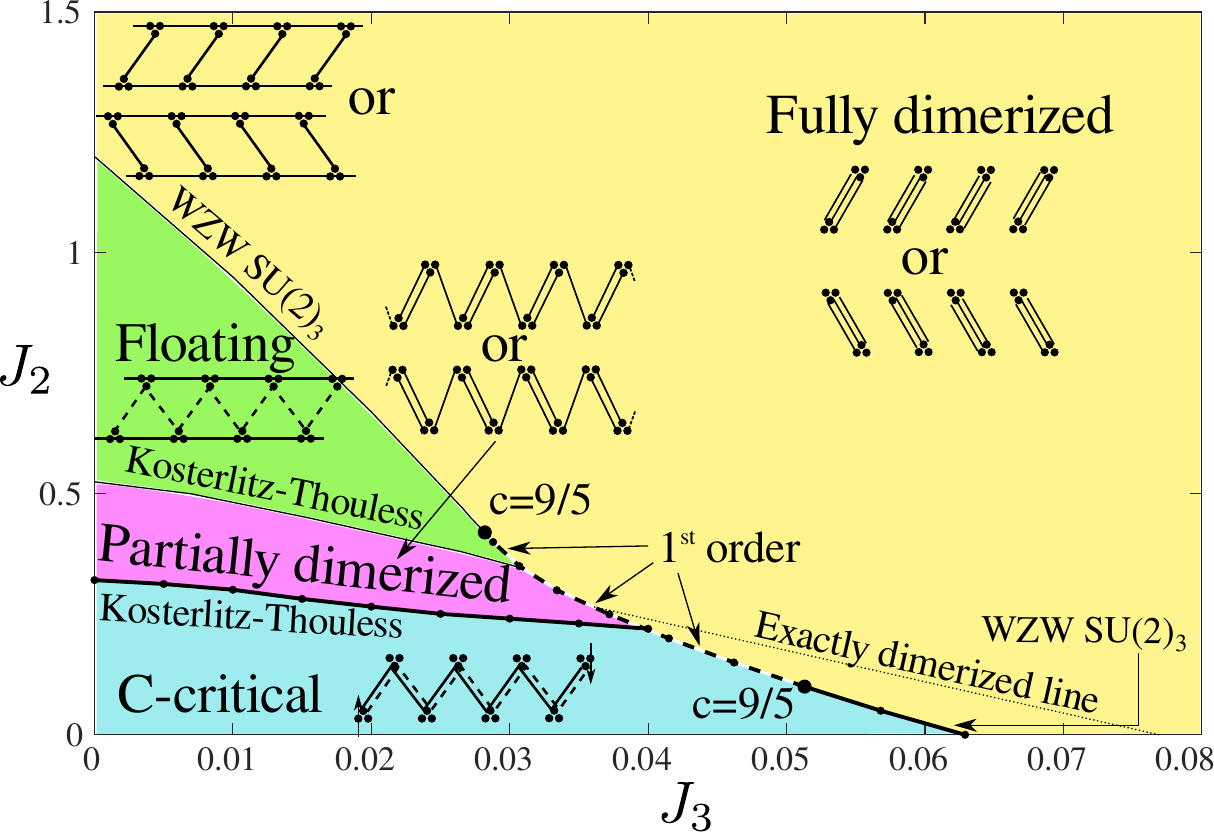}
\caption{Phase diagram of the $S=3/2$ chain with next-nearest-neighbor $J_2$ and three-site interactions $J_3$. Both partially and fully dimerized phases are gapped and spontaneously break the translation symmetry. The fully dimerized state is an exact ground-state along the dotted line. Both the c-critical and the floating phases are characterized by a gapless spectrum and algebraically decaying correlations, however, by contrast to the commensurate-critical phase, the correlations in the floating phase are incommensurate with the lattice. The fully dimerized phase is separated from both the critical and the floating phases by a continuous WZW SU(2)$_3$ transition along the solid line and by a first order transition along the dashed one. The partially dimerized phase is separated from both the floating and critical phases by 
Kosterlitz-Thouless critical lines. The transition between the partially and fully dimerized phases is always first order. The precise location of the boundaries of the floating phase is not known, thin black lines are just indicative.
}
\label{fig:pd_s15}
\end{figure}

Our numerical results are summarized in the phase diagram of Fig.~\ref{fig:pd_s15}. 
It contains two dimerized phases - partially and fully dimerized; and two critical phases with commensurate and incommensurate correlations. 
The dimerized phases can be schematically illustrated using a valence bond singlet (VBS) representation as shown in sketches of Fig.~\ref{fig:pd_s15}. The fully dimerized phase corresponds to three valence-bonds on every other $J_1$ bond, while the partially dimerized phase corresponds to alternating one and two valence bonds.

The conventional or commensurate (c-) critical phase is similar to that of the Heisenberg spin-1/2 chain. It is stabilized when both $J_2$ and $J_3$ couplings are small. The dominant wave-vector of the spin-spin correlations is $q=\pi$. This critical phase can be  described by the WZW SU(2)$_1$ conformal field theory\cite{affleck_haldane}. 
In terms of VBS singlets, this phase can be visualized as one valence bond per $J_1$ bonds, and on top of that one valence bond that resonates between two neighboring bonds (shown schematically with a dashed line).

By contrast, inside the critical phase stabilized at larger values of $J_2$, the wave-vector $q$ is not locked and changes within the phase. Following the classification by Bak\cite{Bak_1982} we will refer to this critical phase as a floating phase. As in the previous case, the underlying critical theory is WZW SU(2)$_1$ topped with incommensurate oscillations, that, in particular, affect the boundary conditions. In terms of VBS singlets, the floating phase can be viewed as a sequence of different domains. The size of the domains, i.e. the period, changes with the wave-vector $q$.

The transition between the c-critical and partially dimerized phases is in the Kosterlitz-Thouless\cite{Kosterlitz}(KT) universality class, in agreement with the previous study of  the $J_1-J_2$ model\cite{roth}.
Both the c-critical phase and the KT critical line are described by WZW SU(2)$_{k=1}$ theory, however, in complete analogy with the critical $J_1-J_2$ spin-1/2 chain, they can be distinguished by logarithmic corrections. Due to the presence of marginal operators the logarithmic corrections appear on top of all finite-size scaling inside the critical phase. By contrast, at the KT critical line the coupling constant of the marginal operator vanishes, and the log-corrections disappear. In Sec.~\ref{sec:KTtransition_s15} we will explain how one can exploit the log-corrections to determine the location of the KT critical line.

The transition between the critical and fully dimerized phases is continuous in the WZW SU(2)$_{k=3}$ universality below and up to the end point and first order beyond it. The end point is located around $J_2\approx 0.1$ and $J_3\approx0.05128$. As has been shown for the spin-1 chain \cite{j1j2j3_short} 
the switch from continuous WZW SU(2)$_k$ to first order transition is induced by the change of sign of a marginal operator. 
The logarithmic corrections that are non-zero all along the continuous transition vanish at the end point, where the marginal coupling vanishes. Interestingly, in the spin-3/2 chain the end point is located at $J_2\approx0.1$, which changes very little compared to the corresponding value for the spin-1 chain $J_2=0.12$\cite{j1j2j3_short}.

The first order transition line continues towards small $J_3$ and separates the fully dimerized from the partially dimerized phase.  
 Then the line continues as a first order transition between the floating and the fully dimerized phases until around $J_2\approx 0.42$ where it eventually turns again into a continuous WZW SU(2)$_3$ critical line. This is remarkable since, to the best of our knowledge, neither a first order transition nor a higher level $k>1$ WZW critical lines has been previously reported at the boundary of a floating phase. 

\begin{figure}[t]
\centering 
\includegraphics[width=0.49\textwidth]{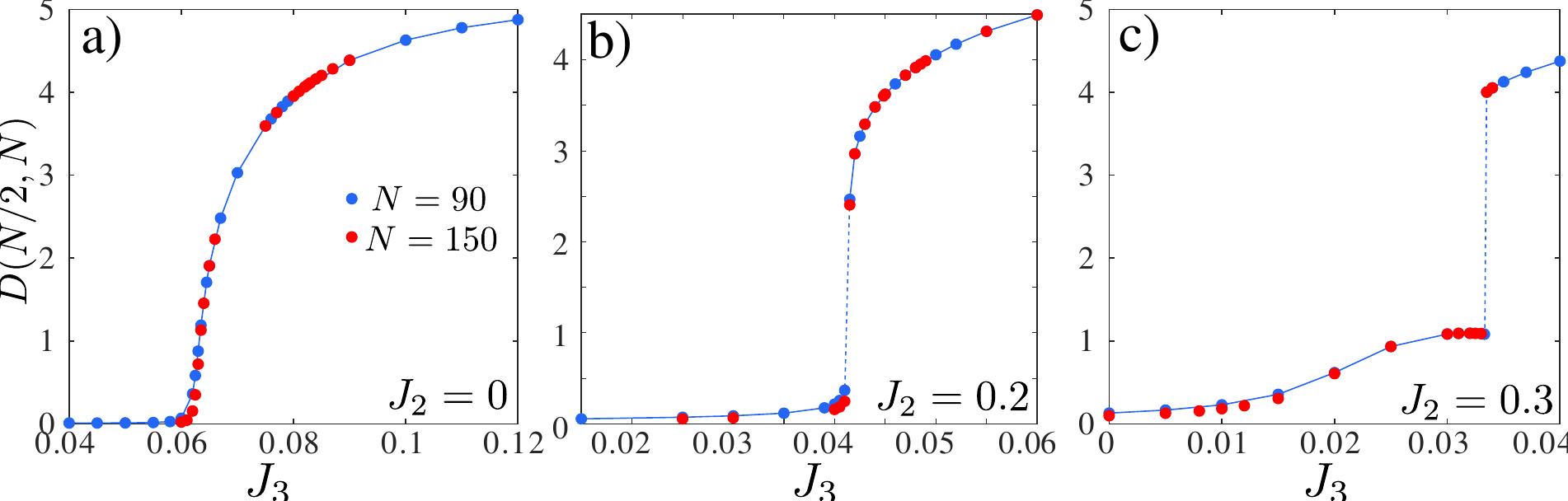}
\caption{Middle chain dimerization for $N=90$ (blue) and $N=150$ (red) across different transitions. (a) Continuous growth of dimerization across the WZW SU(2)$_3$ critical line. (b) Finite jump in dimerization across the first order phase transition from the c-critical phase to the fully dimerized phase. (c) Continuous change of the finite-size dimerization from the  non-dimerized critical phase to the partially dimerized phase across a Kosterlitz-Thouless transition (the critical line goes through $J_2=0.3$, $J_3\approx0.008$) and finite-jump of dimerization across the first order transition from the partially to the fully dimerized phase around $J_3\approx0.033$. In the critical phase, the small value of the dimerization is a finite-size effect.}
\label{fig:dim_ex_s15}
\end{figure}

In order to distinguish non-dimerized and dimerized phases we use the dimerization  $D(j,N)=|\langle \vec S_j\cdot\vec S_{j+1} \rangle - \langle \vec S_{j-1}\cdot\vec S_j \rangle|$ as an order parameter to probe numerically the phase diagram.  Fig.~\ref{fig:dim_ex_s15} shows examples of the middle chain dimerization $D(N/2,N)$ as a function of $J_3$ for three different values of $J_2$. The dimerization changes continuously for $J_2=0$ (Fig.~\ref{fig:dim_ex_s15} (a)), in agreement with a continuous WZW SU(2)$_3$ transition. A finite jump in dimerization as in Fig.~\ref{fig:dim_ex_s15}(b) for $J_2=0.2$ implies a first order phase transition. Both transitions occur from non-dimerized to fully dimerized phases.

For $J_2=0.3$, the dimerization is very small up to the transition from the critical to the partially dimerized phase beyond which it increases up to a value approximately equal to 1. 
 The Kosterlitz-Thouless transition between the two occurs around $J_3\approx0.008$. Under further increase of the $J_3$ coupling, the dimerization jumps abruptly to approximately $D(N/2,N)\approx 4$, indicating a first order phase transition to the fully dimerized phase.

\begin{figure}[h!]
\centering 
\includegraphics[width=0.48\textwidth]{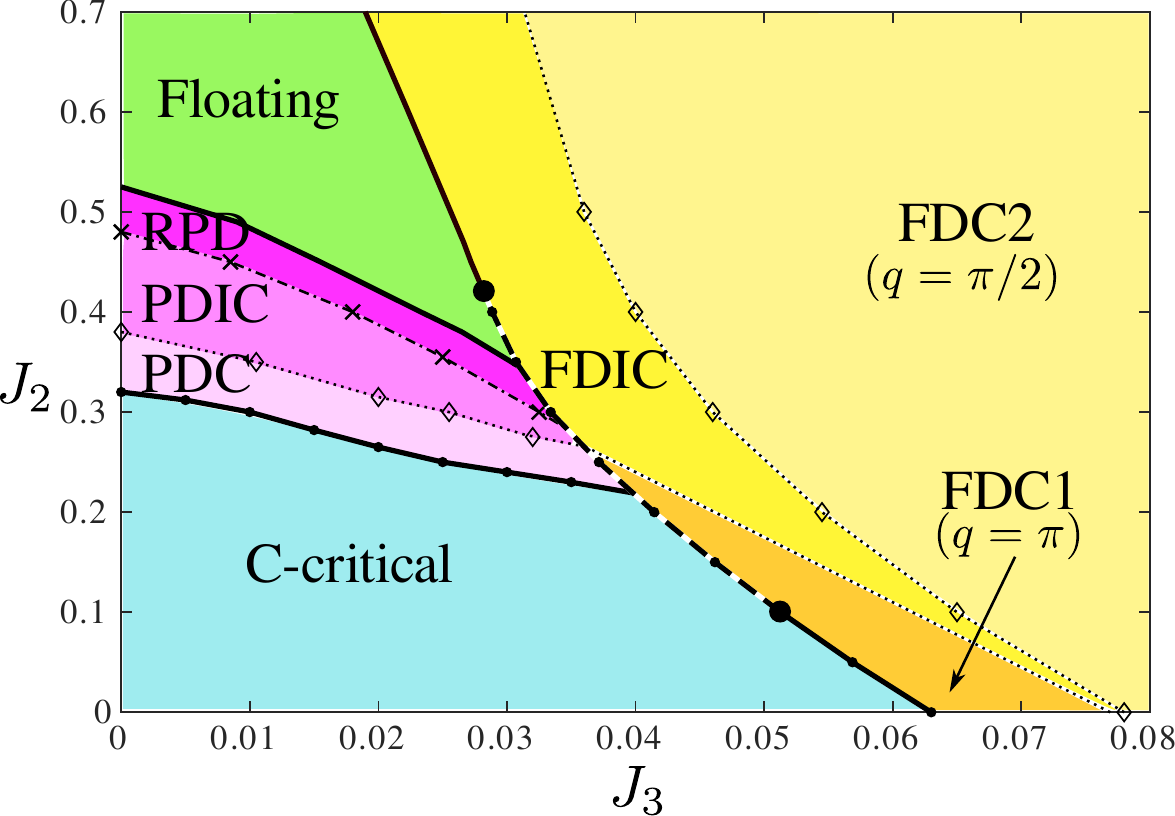}
\caption{Extended phase diagram of the spin-3/2 $J_1-J_2-J_3$ model.
The dotted lines correspond to three disorder lines, one of which coincides with the exactly solvable line given by Eq.\ref{nnn_dimerization}. There are three regions in the fully dimerized (FD) phase that can be distinguished by short-range order: commensurate (C) with wave-vector $q=\pi$ below the exactly dimerized line; incommensurate with $\pi/2<q<\pi$ (FDIC) above the exact line and not too far from the phase boundary and commensurate with $q=\pi/2$ deep inside the fully dimerized phase. The partially dimerized (PD) phase is commensurate below the disorder line and incommensurate above it. Non-protected spin-1/2 edge states are present in the partially dimerized phase in PDC and PDIC and disappear beyond the dash-dotted line. This causes a reorientation of the dimers (RPD) in finite-size chains. Short-range correlations remain incommensurate. The c-critical phase is always commensurate with $q=\pi$. The floating phase is always incommensurate.
}
\label{fig:pd_more}
\end{figure}

On top of long-range dimerization one can distinguish regions of fully and partially dimerized phases by short-range order as shown in Fig.~\ref{fig:pd_more}. Besides, in the partially dimerized phase certain regions can be distinguished by emergent spin-1/2 edge states. Below we provide a short description of each sub-phase.

{\bf Fully dimerized phase:}
\begin{itemize}
  \item {\bf FDC1:} Real-space correlations are commensurate with wave-vector $q=\pi$.
  \item {\bf FDIC:} Real-space correlations are incommensurate with wave-vector $\pi/2<q<\pi$
  \item {\bf FDC2:} Real-space correlations are commensurate with wave-vector $q=\pi/2$.
\end{itemize}

As in the case of spin-1 chain\cite{j1j2j3_long}, the disorder line that separates FDC1 from FDIC coincides with the exact line given by Eq.\ref{nnn_dimerization}.

{\bf Partially dimerized phase:}
\begin{itemize}
  \item {\bf PDC:} Real-space correlations are commensurate with wave-vector $q=\pi$. On finite-size chains there are two spin-1/2 edge states that couple to each other and form a singlet (triplet) ground-state and triplet (singlet) excited state when the entire chain contains an even (odd) number of sites. The energy splitting between these in-gap states is exponentially small with the system size. 
  \item {\bf PDIC:} Real-space correlations are commensurate with wave-vector $q<\pi$. Spin-1/2 edge states are present, but depending on the wave-vector $q$ and the distance between the edges $L$ they can eventually become completely decoupled from each other, which leads to exact zero modes - level crossing between singlet and triplet in-gap states.
  \item {\bf RPD:} As in the PDIC, real-space correlations are commensurate with wave-vector $q<\pi$. Edge states disappear, which leads to the re-orientation of the dimers in finite-size chains: the pattern with (2,1) VBS singlets changes to pattern (1,2).
\end{itemize}


\section{Transition between the c-critical and fully dimerized phases}
\label{sec:CCtoFD}

\begin{figure}[h!]
\centering 
\includegraphics[width=0.5\textwidth]{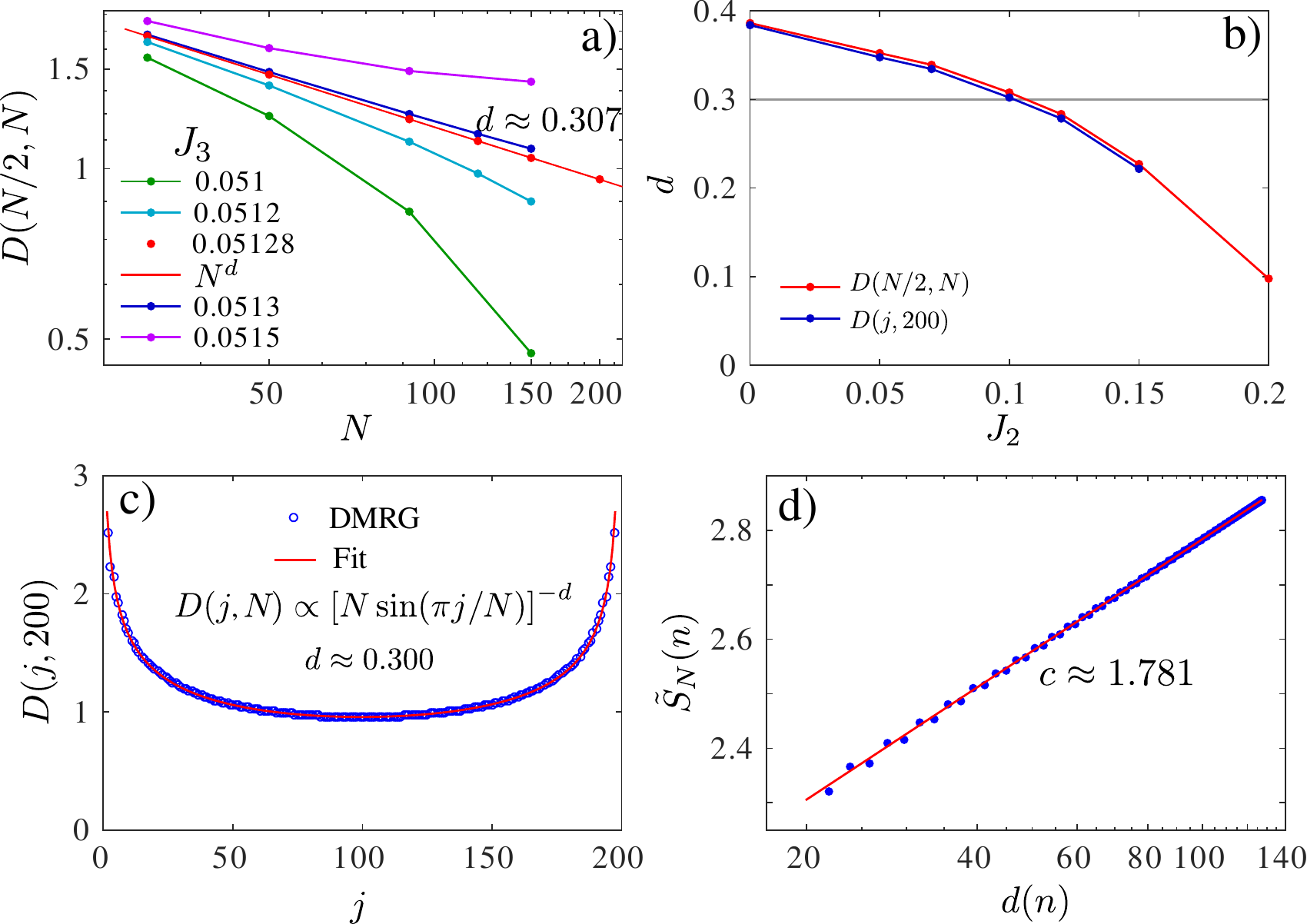}
\caption{(a) Log-log plot of the middle-chain dimerization $D(N/2,N)$ as a function of the number of sites N for $J_2=0.1$ and different parameters $J_3$ around the critical value.  The linear curve corresponds to the critical point and the slope gives the critical exponent $d\approx 0.307$ in good agreement with the CFT prediction $3/10$ for WZW SU(2)$_{k=3}$ critical theory. (b) Apparent critical exponent along the $SU(2)_3$ critical line as a function of $J_2$. Red circles: from the slope of the log-log plot $D(N/2,N)$ as a function of $N$ for the value of $J_3$ for which it is linear. Blue circles: from fitting $D(j,200)$. The black line is the theoretical value of the exponent, $3/10$. Thus the end point is located at $J_2=0.1$ and $J_3=0.05128$. (c) Scaling of the dimerization parameter $D(j,N)$ along the chain with $N=200$ sites at the $SU(2)_3$ critical end point fitted to Eq.\ref{eq:wzwfried_general}. The extracted exponent is in excellent agreement with $d=3/10$. (d) Entanglement entropy at the end point for $N=200$ after removing the Friedel oscillations with weight $\zeta\approx -1.3$. The central charge obtained from the fit to Calabrese-Cardy formula\cite{CalabreseCardy} $c\approx1.781$ agrees within $2\%$ with the CFT predictions $9/5$.}
\label{fig:s15_dimeri_scaling}
\end{figure}

In order to determine the precise location of the critical line between the non-dimerized c-critical phase and the fully dimerized one, we looked at the finite-size scaling of the middle-chain dimerization $D(N/2,N)$. For each fixed value of $J_2$ we compute the dimerization as a function of system size for several values of $J_3$. Then the critical line is associated with a straight line (the separatrix) in a log-log plot of the dimerization $D(N/2,N)$ as a function of the chain length $N$ (see Fig.~\ref{fig:s15_dimeri_scaling}(a)). 

Based on numerical calculations of the central charge,  Michaud et al.\cite{michaud2} have shown that for $J_2=0$ the corresponding transition is in the WZW SU(2)$_3$ universality class. 
Conformal field theory (CFT) predicts for WZW SU(2)$_{k=3}$ the scaling dimension of the dimerization operator to be $d=3/[2(2+k)]=3/10$. The slope of the separatrix gives an `apparent' critical exponent, which is different from $d=3/10$ due to logarithmic corrections. At the end point however, the coupling constant  of the marginal operator vanishes, and the logarithmic corrections disappear. So this is the only point along the transition where the critical exponents can be accurately extracted from finite sizes.  By keeping track of the apparent critical exponent along the critical line, we find that it crosses the line $d=3/10$ at $J_2\approx 0.10$ and $J_3\approx0.05128$ as presented in Fig.~\ref{fig:s15_dimeri_scaling}(b).

It turns out that the fully dimerized phase first appears at the edges of an open chain and therefore free boundary conditions in the spin-3/2 chain correspond to fixed boundary conditions in CFT. A similar effect has been previously reported for the spin-1 $J_1-J_2-J_3$ chain\cite{j1j2j3_short}. Then, according to the boundary CFT, the dimerization in the finite-size chain at the critical line scales (up to logarithmic corrections) as:
\begin{equation}
D(j,N)\propto 1/[(N/\pi) \sin (\pi j/N)]^{d}
\label{eq:wzwfried_general}
\end{equation}
where $j$ is the position index, and the critical exponent is $d=3/10$ for WZW SU(2)$_{k=3}$. This effect is known as Friedel oscillations. An example of the scaling of the dimerization along a finite chain is shown in Fig.~\ref{fig:s15_dimeri_scaling}(c). The critical exponents extracted along the transition line are summarized in Fig.~\ref{fig:s15_dimeri_scaling}(b) and are in a perfect agreement with those extracted from the finite-size scaling of the middle-chain dimerization $D(N/2,N)$.

We extract the central charge numerically from the finite-size scaling of the entanglement entropy in an open chain:
\begin{equation}
\tilde{S}_N(n)=\frac{c}{6}\ln d(n)+\zeta \langle {\bf S}_n{\bf S}_{n+1} \rangle+s_1+\ln g,
\label{eq:calabrese_cardy_obc_corrected_s15}
\end{equation}
where $d=\frac{2N}{\pi}\sin\left(\frac{\pi n}{N}\right)$ is the conformal distance  and $\zeta$ is a non-universal constant introduced in order to suppress Friedel oscillations. Fig.~\ref{fig:s15_dimeri_scaling}(d) provides an example of a fit of the reduced entanglement entropy $\tilde{S}_N(n)$ with Eq.\ref{eq:calabrese_cardy_obc_corrected_s15}.
The values of the central charge along the continuous part of the transition always agree within $3\%$ with the CFT prediction $c=9/5$ for the critical WZW SU(2)$_{3}$ theory.

\begin{figure}[h!]
\centering 
\includegraphics[width=0.48\textwidth]{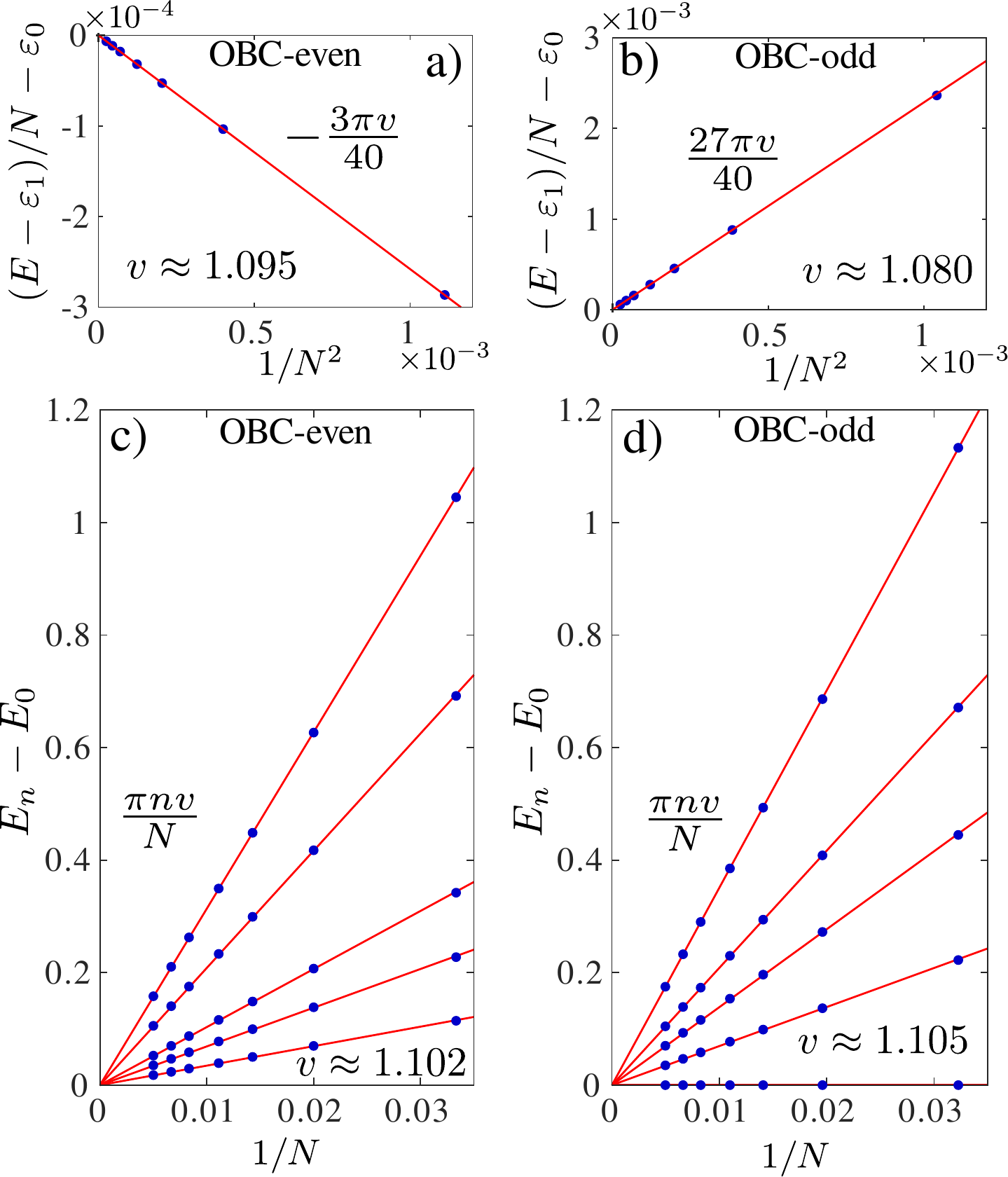}
\caption[Finite-size scaling of the ground-state and excitation energies at the WZW SU(2)$_3$ end point]{Ground state and excitation energy at $J_2=0.1$ and $J_3=0.05128$, on the critical line between the critical and the fully dimerized phases. (a) and (b): Linear scaling of the ground state energy per site with $1/N^2$ after subtracting $\varepsilon_0$ and $\varepsilon_1$ in open chains with (a) even and (b) odd numbers of sites $N$. (c) and (d): Energy gap between the ground state and the lowest energies in different sectors of $S_z^\mathrm{tot}=1,...,5$ for even and $S_z^\mathrm{tot}=5/2,...,11/2$ for odd (blue circles) as a function of $1/N$ for even and odd numbers of sites. Red lines are CFT predictions for $j=0$ and $j=3/2$ towers with the velocities extracted from the first lowest excitation level and indicated in each panel. }
\label{fig:tow_s15}
\end{figure}

For any conformally invariant boundary condition, the ground state scales with the system size as
\begin{equation}
E=\varepsilon_0 N+\varepsilon_1+\frac{\pi v}{N}\left(-\frac{c}{24}+x\right),
\end{equation}
where $\varepsilon_0$ and $\varepsilon_1$ are non-universal constants, $c$ is the central charge and $x$ is the scaling dimension of the corresponding primary field.
For the $SU(2)_{k=3}$ WZW model there are 4 conformal towers labeled by the spin of the lowest energy states, $j=0$, $1/2$, $1$ and $3/2$. The scaling dimension of the corresponding operator is given by $x=j(j+1)/(2+k)$. Chains with an even number of sites have a singlet ground-state and are thus described by the conformal tower $j=0$ with scaling dimension $x=0$. By contrast, the ground state of a chain with an odd number of sites belongs to the conformal  tower with the largest $j=3/2$ with scaling dimension $x=3/4$. Thus the ground state energies of an open chain with even or odd numbers of sites scale as:
\begin{eqnarray}
E_\mathrm{even}=\varepsilon_0 N+\varepsilon_1-\frac{3\pi v}{40N},\\
E_\mathrm{odd}=\varepsilon_0 N+\varepsilon_1+\frac{27\pi v}{40N}.
\end{eqnarray}
Examples of finite-size scaling of the ground-state energy for even and odd numbers of sites are shown in Fig.~\ref{fig:tow_s15}(a) and (b).

We have extracted several excited states by computing the lowest states within different symmetry sectors of total magnetization $0\leq S^z_\mathrm{tot}\leq 5$.
In order to construct WZW SU(2)$_{k=3}$ conformal towers we have closely followed Ref.~\onlinecite{AffleckGepner}. Since we are interested only in the lowest state for different values of the total spin $s$, the energy level that corresponds to this state is defined by an integer $n$ that satisfies:
\begin{equation}
 \frac{j^2-S^2}{k}\leq n<\frac{j^2-S^2+k}{k}
\label{eq:conftower}
\end{equation}
 The results for $j=0$ and $j=3/2$  WZW SU(2)$_{3}$ conformal towers are summarized in Table \ref{tb:wzwsu2k3}. For the $j=3/2$ tower, the ground-state is in the sector with $S_\mathrm{tot}=3/2$, and it appears as the lowest state in the two sectors of total magnetization $S^z_\mathrm{tot}=1/2$ and $3/2$. 

\begin{table}[h!]
\centering
\begin{tabular}{|c|c|c|c|c|c|c|}
\multicolumn{7}{c}{j=0}\\
\hline 
$s$&0&1&2&3&4&5\\
\hline 
$(E-E_0)N/ \pi v$ &0&1&2&3&6&9\\
\hline 
\multicolumn{7}{c}{j=3/2}\\
\hline 
$s$&1/2&3/2&5/2&7/2&9/2&11/2\\
\hline 
$(E-E_0)N/ \pi v$, & 2&0&2&4&6&10\\
\hline 
\end{tabular}
\caption{Lowest excitation energy with spin $s$ for both $j=0$ and $j=3/2$ WZW SU(2)$_{3}$ conformal towers.}
\label{tb:wzwsu2k3}
\end{table}

The conformal towers obtained numerically for both even and odd numbers of sites are shown in Fig.~\ref{fig:tow_s15}(c) and (d) and summarized in Table \ref{tb:resconftow_wzwsu2k3}.

\begin{table} 
\centering
\begin{tabular}{l||r|r}
 &  &DMRG: $J_2=0.1$\\
Energy level& CFT SU(2)$_3$&$J_3=0.05128$\\
\hline \hline
OBC, Even, GS $S_z^\mathrm{tot}=0$ &-3/40& -3/40\\
\hline 
OBC, Even, GS $S_z^\mathrm{tot}=1$  &1& 1.0065\\
\hline
OBC, Even,  GS  $S_z^\mathrm{tot}=2$ &2& 2.0003\\ 
\hline
OBC, Even,  GS  $S_z^\mathrm{tot}=3$ &3& 2.9999\\
\hline 
OBC, Even,  GS  $S_z^\mathrm{tot}=4$ &6& 6.057\\
\hline 
OBC, Even, g GS  $S_z^\mathrm{tot}=5$ &9& 9.12\\
\hline \hline 
OBC, Odd,  GS  $S_z^\mathrm{tot}=3/2$  &27/40& \\
&=0.675 & 0.666\\
\hline
OBC, Odd,  GS  $S_z^\mathrm{tot}=5/2$ &2& 2.018\\ 
\hline
OBC, Odd,  GS  $S_z^\mathrm{tot}=7/2$ &4& 4.028\\
\hline 
OBC, Odd,  GS  $S_z^\mathrm{tot}=9/2$ &6& 6.058\\
\hline 
OBC, Odd,  GS  $S_z^\mathrm{tot}=11/2$ &10& 10.19
\end{tabular}
\caption{ Energy levels at the $SU(2)_3$ critical point. The ground state for $N$ even $S_z^\mathrm{tot}=0$ and  odd $S_z^\mathrm{tot}=3/2$ refers to the $1/N$ term in the ground state energy.  For the rest, the gap above the ground state is given. The results are in units of $\pi v/N$ with $v=1.095$.} 
\label{tb:resconftow_wzwsu2k3}
\end{table}

\begin{figure}[h!]
\centering 
\includegraphics[width=0.49\textwidth]{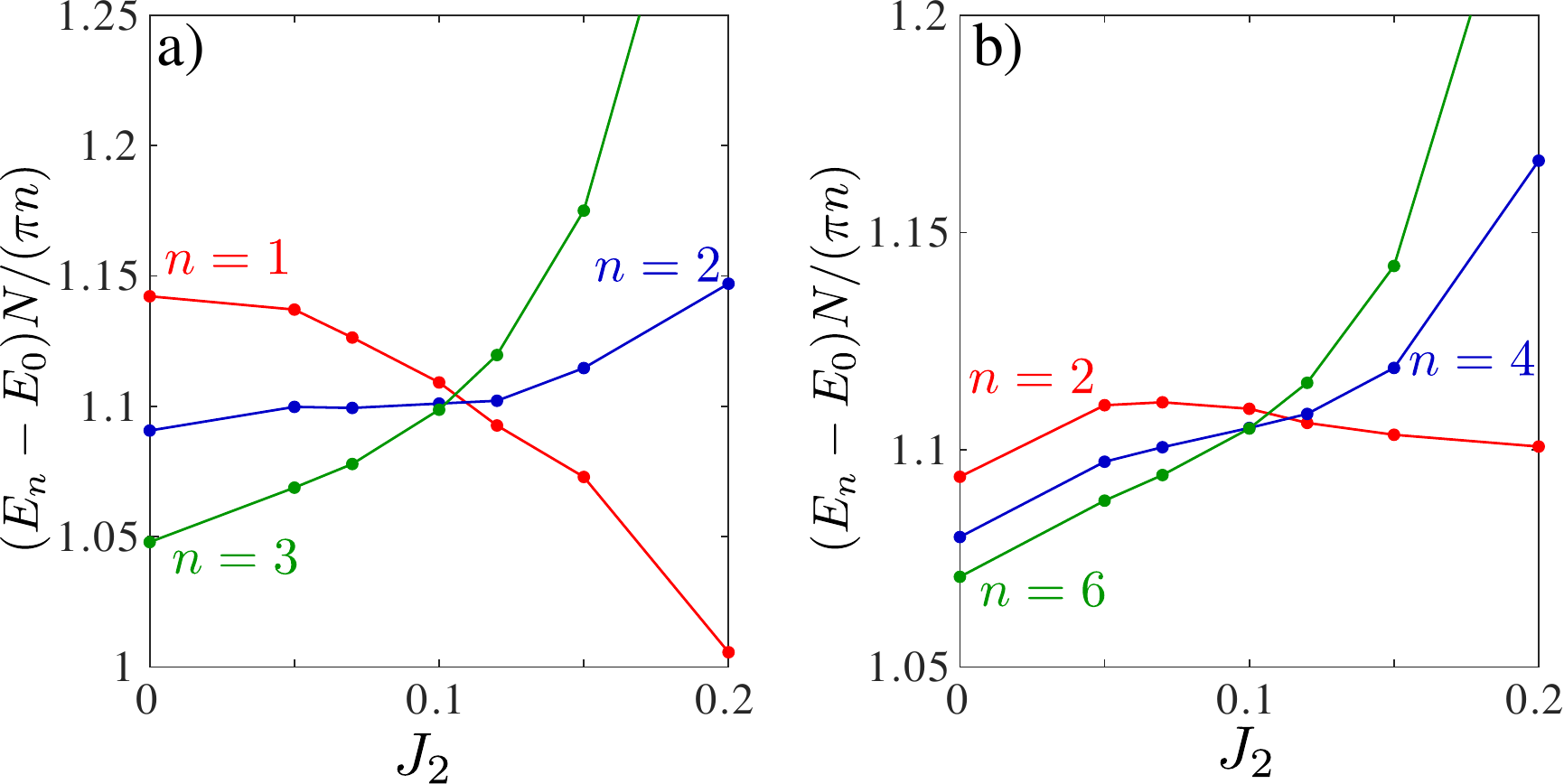}
\caption{Velocity along the critical line between the critical and the fully dimerized phases extracted from the gap between the $n^\mathrm{th}$ energy level and the ground state for (a) $N=50$  and (b) $N=51$}
\label{fig:velocity_s15}
\end{figure}

Finally, in order to prove that the pair of parameters $J_2=0.1$ and $J_3=0.05128$ indeed corresponds to the end point, we show that the conformal tower is destroyed by moving along the critical line away from the end point.
Following the procedure established in Ref.~\onlinecite{j1j2j3_short}, we have plotted the velocities extracted from three different excitation levels $n$ according to $v_n=(E_n-E_0)N/(\pi n)$  (Fig.~\ref{fig:velocity_s15}). At the end point, all velocities are the same, implying that the conformal tower is restored. This occurs around $J_2=0.1$, in agreement with the value determined from the critical exponent. 
Due to logarithmic corrections the velocities split but remain relatively close to each other along the continuous transition. Above the end point however the spitting of $(E_n-E_0)N/(\pi n)$ is much faster, in agreement with the first order transition with a very different structure of the spectrum.

In order to characterize the phase transition beyond the end point we have looked at the dimerization and the ground-state energy. Both quantities were computed in the middle of fairly long chains with $N=200$ and $N=400$ sites to reduce the impact from the finite-size effects and provide an estimate of their values in the thermodynamic limit. The energy per bond $\varepsilon_N$ is defined by:

$$\varepsilon_{N}=\varepsilon_1+\varepsilon_2+\varepsilon_3,$$
where
$$\varepsilon_1=\frac{J_1}{2}\langle{\bf S}_{i-1}\cdot {\bf S}_i+{\bf S}_i\cdot{\bf S}_{i+1}\rangle,$$
$$\varepsilon_2=J_2\langle{\bf S}_{i-1}\cdot {\bf S}_{i+1}\rangle,$$
$$\varepsilon_3=J_3\langle({\bf S}_{i-1}\cdot {\bf S}_i)({\bf S}_i\cdot {\bf S}_{i+1})+{\mathrm h.c.}\rangle,$$
where $(i,i+1)$ is the central bond.

\begin{figure}[h!]
\centering 
\includegraphics[width=0.49\textwidth]{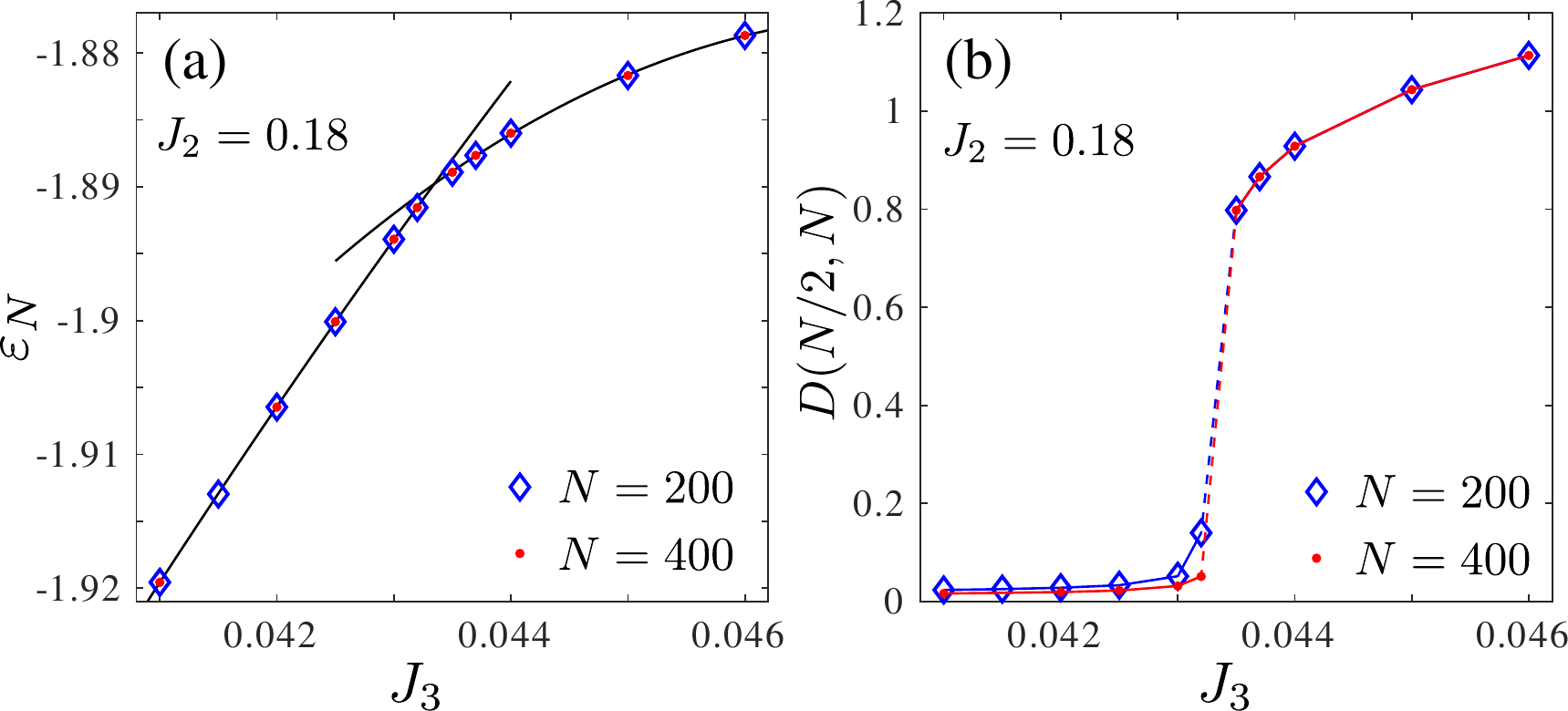}
\caption{(a) Kink in the energy per site plotted as a function of $J_3$ for $J_2=0.18$ above the end point. (b) Finite jump in the dimerization as a function of $J_3$ for $J_2=0.18$ in agreement with the first order phase transition.}
\label{fig:firstorder}
\end{figure}

Beyond the end point, for $J_2=0.18$, we detect a kink in the ground-state energy $\varepsilon_N$ as shown in Fig.~\ref{fig:firstorder}(a). In the vicinity of the transition the energy increases monotonously with $J_3$; thus in order to see the change of the slope we have to look at the narrow window around the phase transition. 
By extrapolating the numerical data with a second order polynomial (black lines in Fig.~\ref{fig:firstorder}(a)) we find the crossing point of the two fits around $J_3=0.0433$. It is essential that the ground-state energy does not change significantly upon increasing the system size from $N=200$ to $N=400$, so that the edge effects are negligibly small in the middle of the chain for the chosen system sizes.

 We also detect a finite-jump in the dimerization $D(N/2,N)$  between $J_3=0.0432$ and  $J_3=0.0435$ as shown in Fig.~\ref{fig:firstorder}(b). Thus the transition is expected to occur within this interval in agreement with the results obtained from the ground-state energy. 
 
 To summarize, we have provided numerical evidence that the phase transition between the critical and the fully dimerized phases is continuous in the WZW SU$(2)_3$ universality class below and including at the end point and it is first order beyond it. This result is rather surprising since the critical phase is connected to a gapped phase with spontaneously broken symmetry via a first order transition. To the best of our knowledge such a scenario has not yet been observed in the context of one-dimensional spin systems. 
 The conformal field theory explains the appearance of the first order transition by the change of the sign of the marginal coupling constant.


\section{C-critical phase and first Kosterlitz-Thouless transition}
\label{sec:KTtransition_s15}

As pointed out above, the commensurate critical phase that appears at small values of $J_2$ and $J_3$ and the Kosterlitz-Thouless transition to the partially dimerized phase are both characterized by the WZW SU(2)$_1$ critical theory and can be distinguished only by the logarithmic corrections. 

\begin{figure}[h!]
\centering 
\includegraphics[width=0.49\textwidth]{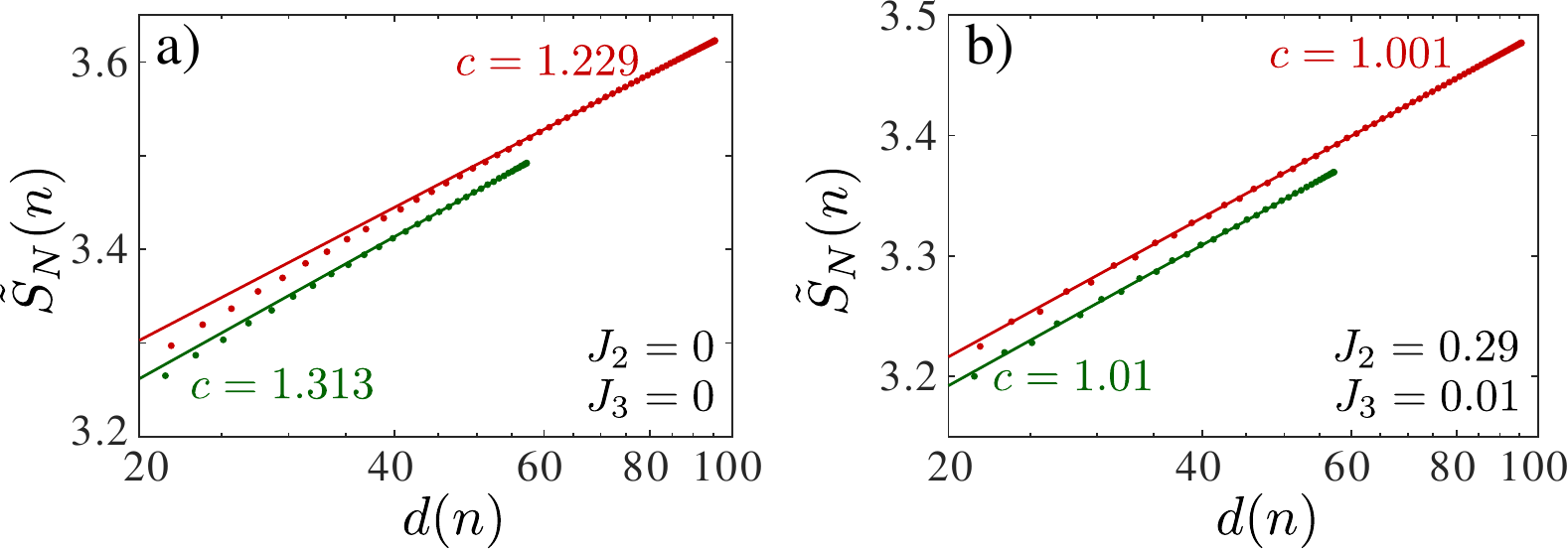}
\caption[Central charge inside the critical phase and at the Kosterlitz-Thouless critical line]{ Extraction of the central charge for open chains with $N=90$ (green) and $N=150$ (red) by fitting the reduced entanglement entropy $\tilde{S}_N(n)$ with the Calabrese-Cardy formula of Eq.\ref{eq:calabrese_cardy_obc_corrected_s15} inside the critical phase (a) far from and (b) close to the Kosterlitz-Thouless transition}
\label{fig:cc_su2k1_s15}
\end{figure}

We have extracted the central charge numerically by fitting the reduced entanglement entropy to Eq.\ref{eq:calabrese_cardy_obc_corrected_s15}. Examples of fits of finite-size results are provided in Fig.~\ref{fig:cc_su2k1_s15}. CFT predicts the central charge $c=1$ for WZW SU(2)$_1$ critical theory.
Due to large logarithmic corrections the central charge extracted from the entanglement entropy in finite-size clusters differs significantly from this prediction deep inside the critical phase, as can be observed in Fig.~\ref{fig:cc_su2k1_s15}(a). By contrast, close to the Kosterlitz-Thouless critical line, the logarithmic corrections are suppressed, and the central charge can be extracted with sufficient accuracy even from relatively small chains Fig.~\ref{fig:cc_su2k1_s15}(b).

\begin{figure}[h!]
\centering 
\includegraphics[width=0.49\textwidth]{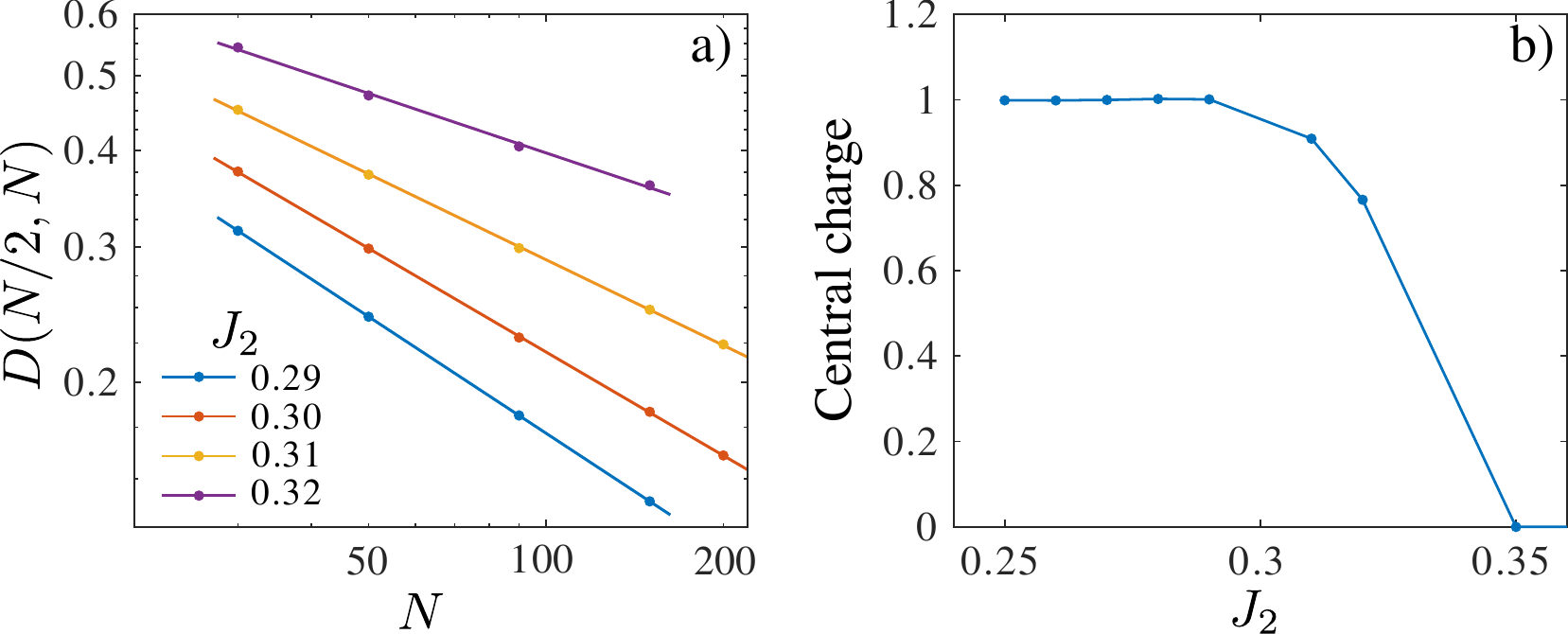}
\caption{ (a) Middle-chain dimerization $D(N/2,N)$ as a function of the system size $N$ for $J_3=0.01$. For finite-size systems the apparent finite-size scaling is linear even above the transition that occur at  $0.29\geq J_2\geq 0.30$. (b) Central charge as a function of $J_2$ for $J_3=0.01$. It is given by $c=1$ inside the critical phase close to the transition and decreases towards zero in the gapped partially dimerized phase. The results are for an open chain with $N=150$  sites.}
\label{fig:location_KT_s15}
\end{figure}

Close to the Kosterlitz-Thouless transition the dimerization decreases almost linearly on a log-log scale on both sides of the transition, and 
locating the  critical line by identifying the separatrix becomes extremely challenging (see Fig.~\ref{fig:location_KT_s15}(a)). 
To locate the phase transition, one can extract the central charge, that differs slightly from $c=1$ inside critical phase close to the transition, but rapidly decreases in the gapped partially dimerized phase, as shown in Fig.~\ref{fig:location_KT_s15}. 

\begin{figure}[h!]
\centering 
\includegraphics[width=0.49\textwidth]{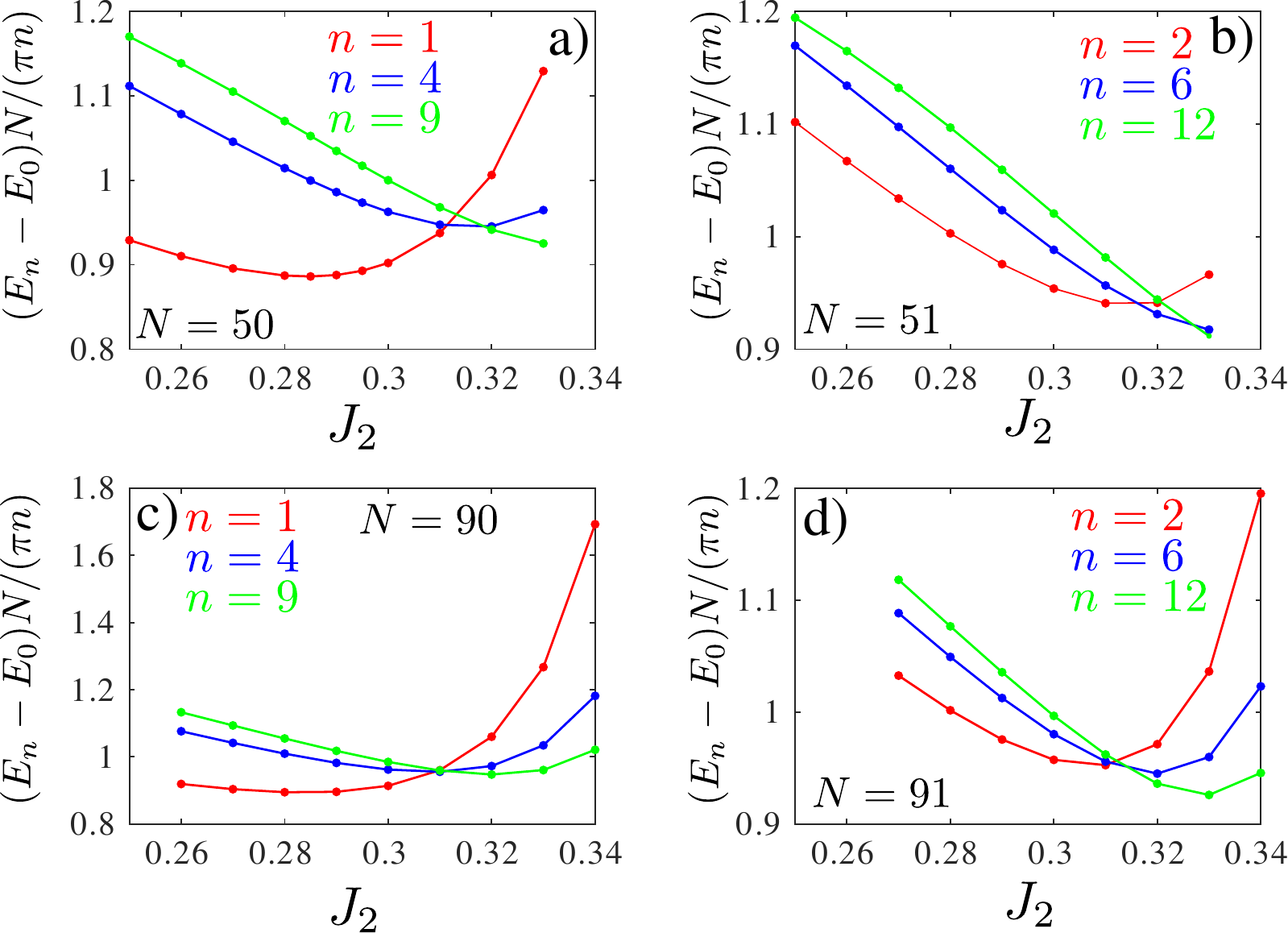}
\caption[Velocities across the Kosterlitz-Thouless transition]{ Velocities across the Kosterlitz-Thouless transition between the critical and partially dimerized phases extracted from the gap between various energy levels and the ground state as a function of $J_2$ for fixed value of $J_3=0.01$ and different system sizes.  }
\label{fig:velocity_KT_s15}
\end{figure}

An alternative way to locate the KT phase transition is based on the effective velocities that can be extracted from the excitation spectrum.
 For the $SU(2)_{k=1}$ WZW model, there are only two conformal towers labeled by the total spin: $j=0$ and $1/2$.
The levels corresponding to the lowest states with different magnetization sectors can be extracted from Eq.~\ref{eq:conftower}. They are summarized in Table \ref{tb:wzwsu2k1}. Due to presence of the low-lying edge states around the Kosterlitz-Thouless transition, the listed states can be approximately found as ground states in the symmetry sector $S^z_\mathrm{tot}=s+1$. The numerical results obtained for $J_3=0.01$ are summarized in Fig.~\ref{fig:velocity_KT_s15}. The crossing point where all velocities are almost the same and therefore the conformal towers are restored is around $J_2\approx 0.31$ and slowly decrease with increasing system size. Thus these results are consistent with our previous estimate of the critical point $J_2\approx0.3$ for $J_3=0.01$. 
 In general one can improve the results by removing or fixing the edge states, which themselves contribute logarithmically to the energy. Here, since we are interested only in the location of the critical line, shifting the sectors by $S^z_\mathrm{tot}=+1$ seems sufficient.

\begin{table}[h!]
\centering
\begin{tabular}{|c|c|c|c|c|c|c|}
\multicolumn{7}{c}{j=0}\\
\hline 
$s$&0&1&2&3&4&5\\
\hline 
$(E-E_0)N/ \pi v$ &0&1&4&9&16&25\\
\hline 
\multicolumn{7}{c}{j=1/2}\\
\hline 
$s$&1/2&3/2&5/2&7/2&9/2&11/2\\
\hline 
$(E-E_0)N/ \pi v$, &0&2&6&12&20&30\\
\hline 
\end{tabular}
\caption{Lowest excitation energy with spin $s$ for both $j=0$ and $j=1/2$ WZW SU(2)$_{1}$ conformal towers.}
\label{tb:wzwsu2k1}
\end{table}


\section{Transition between the two dimerized phases}
\label{sec:dimer_to_dimer}

\begin{figure}[h!]
\centering 
\includegraphics[width=0.47\textwidth]{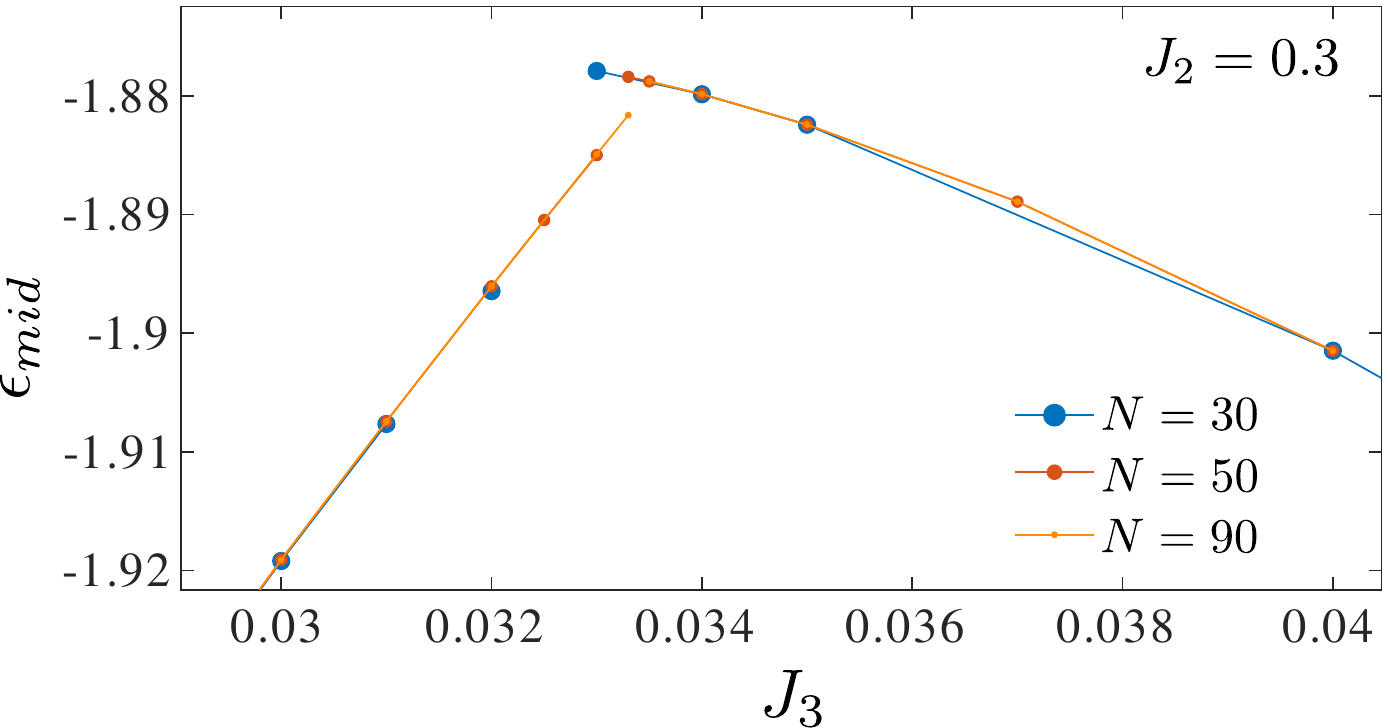}
\caption{ Kink in the energy across the first order phase transition between partially and fully dimerized phases.}
\label{fig:s15_energy_firstorder}
\end{figure}

A phase transition between the partially and the fully dimerized phases occurs for $0.22 \leq  J_2 \leq 0.35$, and it is of first order. This transition can be seen as a pronounced kink in the energy per site $\epsilon_{mid}$ calculated in the middle of the chain. A small hysteresis behavior appears because the dimerization is favored at the open edges (see Fig.~\ref{fig:s15_energy_firstorder}). It decreases with increasing system size. Apart from that, the finite-size effects are very small, and the location of the critical point can be extracted accurately from relatively small clusters.

\begin{figure}[h!]
\centering 
\includegraphics[width=0.47\textwidth]{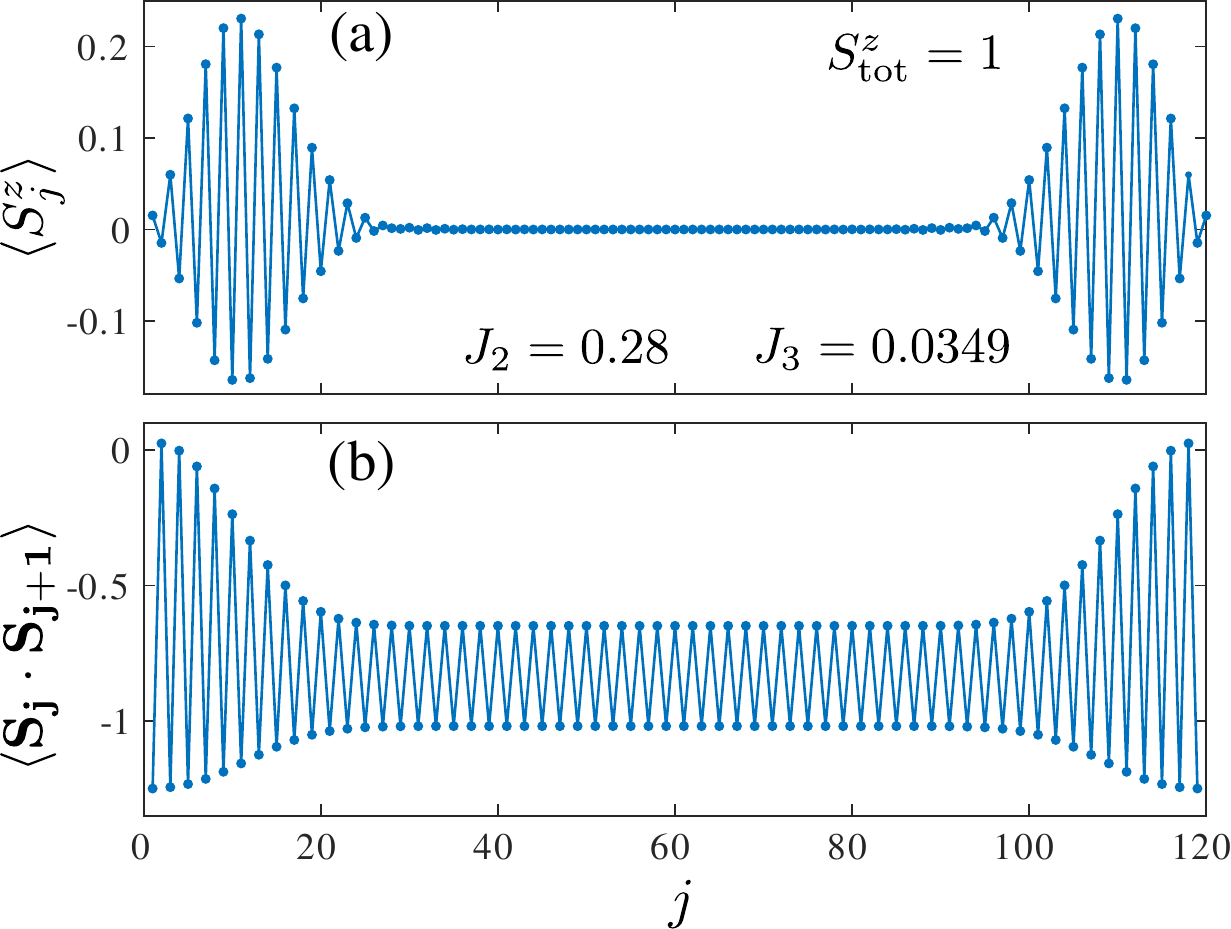}
\caption{(a) Local magnetization and (b) nearest-neighbor correlation profiles for $N=120$ at $J_2=0.28$ and $J_3=0.0349$, a point on the the first order transition line between the partially and fully dimerized states. One can observe the coexistence of domains in the partially and fully dimerized states separated by two domain walls, each of which carries a spin-1/2.}
\label{fig:Solitons}
\end{figure}

The simplest domain wall between the fully and partially dimerized domains carries spin-1/2. One can detect it as a pair of solitons (see Fig.~\ref{fig:Solitons}(a)) in the magnetization profile of a chain with $S^z_\mathrm{tot}=1$ at the transition line between the two phases. From Fig.~\ref{fig:Solitons}(b) one can also conclude that the domains of fully dimerized states are located close to the open edges of the chain, while the domain in the middle is in the partially dimerized state.

\section{Floating phase}
\label{sec:floating}

The numerical investigation of floating phases is always challenging. Since the floating phase is critical and characterized by a divergent correlation length a proper convergence in DMRG can be achieved for relatively small system sizes only. At the same time, the incommensurate wave-vector $q$ has to be compatible with the boundary conditions, either open (and usually spontaneously fixed) or periodic. Therefore the system size should be sufficiently large to resolve the true wave-vector. The closer $q$ is to a commensurate value, the longer the system should be to resolve the difference. Moreover, an increasing next-nearest-neighbor interaction naturally increases the amount of entanglement carried by the $J_2$ bonds, that are superimposed when the system is bipartite in the DMRG.
We keep up to 1500 states and perform up to 7 DMRG sweeps in the two-site routine.

\subsection{Incommensurate correlations}

In order to extract the  wave-vector $q$ we fit the real-space spin-spin correlations
$C_{i,j}=\langle {\bf S}_i \cdot {\bf S}_j \rangle-\langle {\bf S}_i\rangle\cdot\langle{\bf S}_j \rangle$
 to the Ornstein-Zernicke form:

\begin{equation}
  C^\mathrm{OZ}_{i,j}\propto \frac{e^{-|i-j|/\xi}}{\sqrt{|i-j|}}\cos(q|i-j|+\varphi_0),\label{eq:OZ}
\end{equation}
where the correlation length $\xi$, the wave-vector $q$ and the phase shift $\varphi_0$ are fitting parameters. We equally use the same form to fit the correlations inside the critical phase, because the finite length of the chain and the finite MPS bond dimension both induce an effective finite correlation length. We find that the quality of the fit is improved if it is done in two steps. First, we discard the oscillations and fit the main slope of the exponential decay. This allows us to perform a fit in a semi-log scale $\log C(x=|i-j|)\approx c-x/\xi-\log(x)/2$. Second, we define a reduced correlation function
\begin{equation}
  \tilde C_{i,j}=\frac{\sqrt{|i-j|}}{e^{-|i-j|/\xi+c}}C_{i,j} 
\end{equation}
and fit it with a cosine $\tilde C_{i,j}\approx a\cos(q|i-j|+\varphi_0)$.

\begin{figure}[h!]
\centering 
\includegraphics[width=0.47\textwidth]{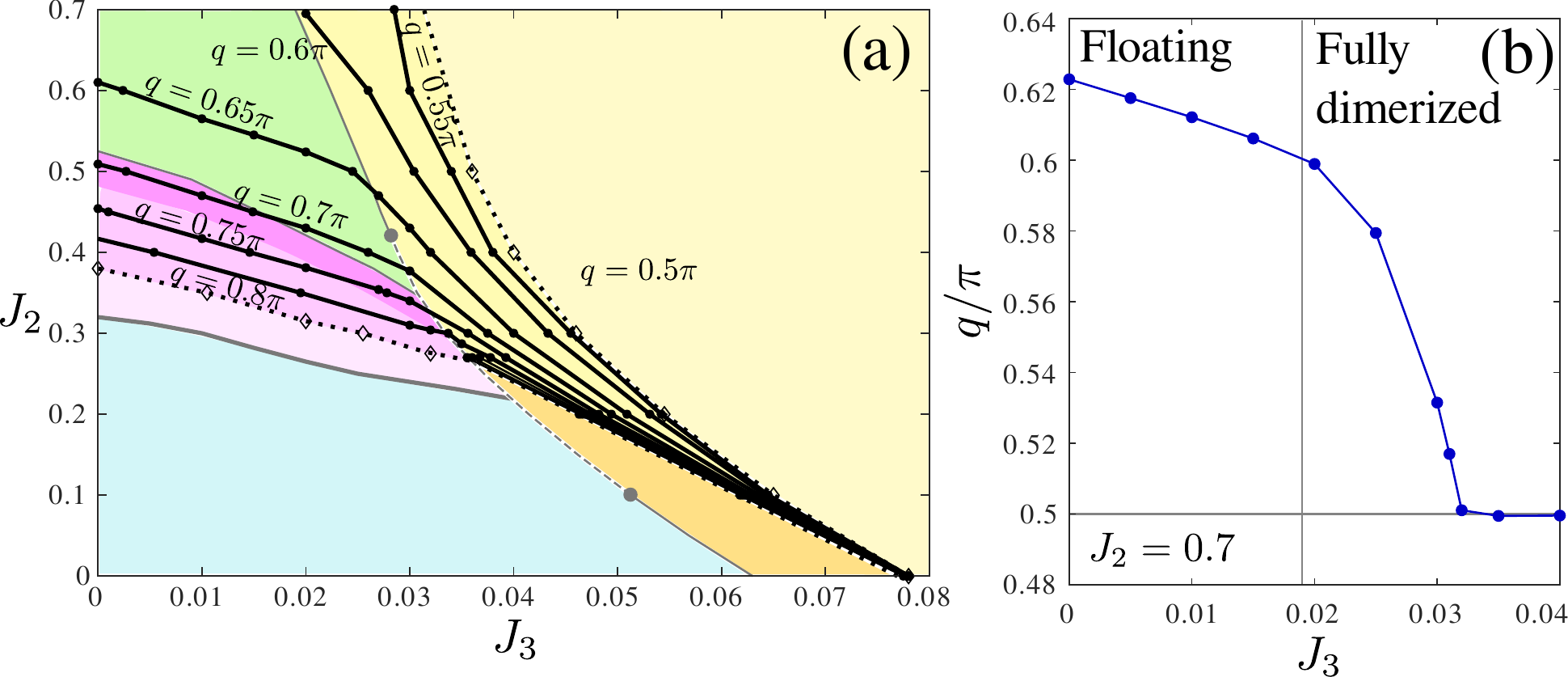}
\caption{(a) Phase diagram with the lines of constant wave-vector $q$ extracted by fitting real-space correlations to the Ornstein-Zernicke form. (b) Dependence of the wave-vector $q$ as a function of $J_3$ for $J_2=0.7$ across a range that overlaps with the floating and fully dimerized phases. The wave-vector $q$ reaches its commensurate value $\pi/2$ at the disorder point $J_3\approx0.032$}
\label{fig:EqualQ}
\end{figure}

Fig.~\ref{fig:EqualQ}(a) summarizes our results and shows three disorder lines and a set of equal-q lines. Close to the exact line in the fully dimerized states the wave-vector changes very fast and the equal-q lines are quite condensed. In the partially dimerized phase, we observe an abrupt change from $q=\pi$ to $q\approx0.83\pi$ at the disorder line. The wave-vector $q$ seems to be locked at this value for a short parameter range not too far from the disorder line. We believe that this plateau is purely a finite-size effect, however a deeper understanding of its nature is beyond the scope of this work. Apart from that, the $q$-vector changes with the parameters $J_2$ and $J_3$ in a continuous and smooth way, in particular inside the floating phase. Fig.~\ref{fig:EqualQ}(b) shows q as a function of $J_3$ for a fixed value of $J_2$. One can clearly see that inside the floating phase the wave-vector is not locked to a particular value, but indeed changes very smoothly with coupling constant.

\begin{figure}[h!]
\centering 
\includegraphics[width=0.47\textwidth]{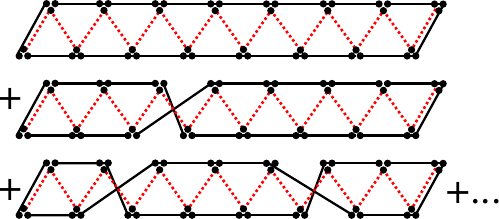}
\caption{VBS sketch of the state in the floating phase that consists of the intertwined Haldane chains (balck solid lines) and a critical spin-1/2 chain (red dashed line).}
\label{fig:incom_sketch}
\end{figure}

In order to understand the nature of the floating phase, let us think of the spin-3/2 chain as a composite system made of a spin-1 chain and a spin-1/2 chain. At sufficiently large value of the next-nearest-neighbor coupling $J_2$, the spin-1 chain is in the next-nearest-neighbor (NNN) Haldane phase with one VBS singlet per $J_2$ bond\cite{kolezhuk_prl}. It has been shown however that this simple picture is only true at infinitely large $J_2$. At finite values, the ground state corresponds to what was called the intertwined Haldane chain\cite{kolezhuk_connectivity}, the periodic twist between the two chains being responsible for the incommensurate short-range correlations. The superposition of this spin-1 state with the remaining critical spin-1/2 sketched in Fig.~\ref{fig:incom_sketch} then naturally leads to an intuitive picture of the floating phase.

\subsection{Dimerization}

\begin{figure}[h!]
\centering 
\includegraphics[width=0.47\textwidth]{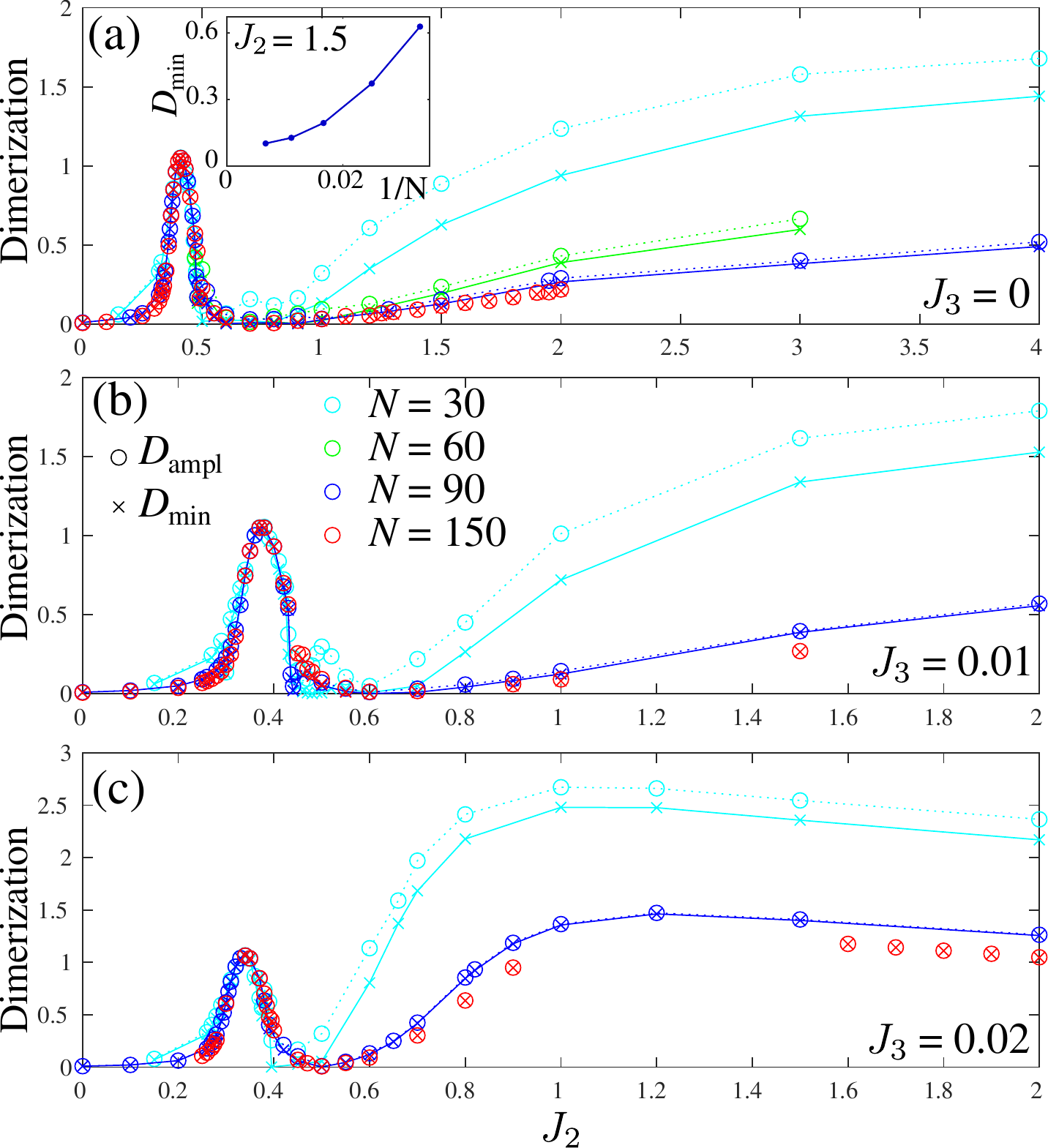}
\caption{Dimerization as a function of $J_2$ for (a) $J_3=0$; (b) $J_3=0.01$; and (c) $J_3=0.02$ and for four different system sizes $N=30,60,90,150$. The first peak of dimerization around $J_2\approx0.3$ corresponds to the partially dimerized phase, while for large $J_2$ the finite dimerization corresponds to the fully dimerized phase. Between these two regimes, the vanishingly small dimerization corresponds to the floating phase. The inset in panel (a) shows the finite-size scaling of the dimerization at $J_3=0$ and $J_2=1.5$.}
\label{fig:Dim_vert}
\end{figure}

Inside the floating phase,  the incommensurability affects also local quantities such as the nearest-neighbor correlations at the center of the chain because the correlation length diverges; therefore the definition of the dimerization $D(N/2,N)$ that we used before is no longer applicable. Instead, we compute the amplitude of the nearest-neighbor correlations:
\begin{equation}
  D_\mathrm{ampl}=\max(\langle{\bf S}_i{\bf S}_{i+1}\rangle)-\min(\langle{\bf S}_i{\bf S}_{i+1}\rangle)
\end{equation}
and the minimum of the local dimerization:
\begin{equation}
  D_\mathrm{min}=\min(\langle{\bf S}_{i-1}{\bf S}_{i}\rangle-\langle{\bf S}_{i}{\bf S}_{i+1}\rangle),
\end{equation} 
 where $N/2-5<i<N/2+5$, i.e. in a small window in the middle of the chain. Inside the dimerized phase (when the system size sufficiently exceeds the correlation length) both definitions give the same result as the middle-chain dimerization $D(N/2,N)$  used before.
In Fig.~\ref{fig:Dim_vert} we show $ D_\mathrm{ampl}$ and $ D_\mathrm{min}$ as a function of the next-nearest-neighbor interaction $J_2$ for three values of $J_3$. In each of the three cases the curve starts with a small finite-size dimerization inside the commensurate critical phase, followed by a pronounced peak with finite dimerization over an extended region that corresponds to the partially dimerized phase. Upon further increase of $J_2$ the dimerization is non monotonous - it decreases and remains extremely small over an extended parameter range; eventually it increases again indicating the entrance to the fully dimerized phase. By analogy with spin-1/2 we  expect the dimerization to decrease  at large $J_2$; our results for $J_3=0.02$ support this scenario.

\subsection{Second Kosterlitz-Thouless transition}

Finding the location of the Kosterlitz-Thouless phase transition between the partially dimerized and the floating phase is extremely challenging because of the incommensurate correlations on both sides of the transition. In particular, for any finite size chain, open edges with stronger dimerization correspond to  conformally invariant boundary conditions  only at particular values of the wave-vector $q$, and therefore only at specific points in the phase diagram.

\begin{figure}[h!]
\centering 
\includegraphics[width=0.47\textwidth]{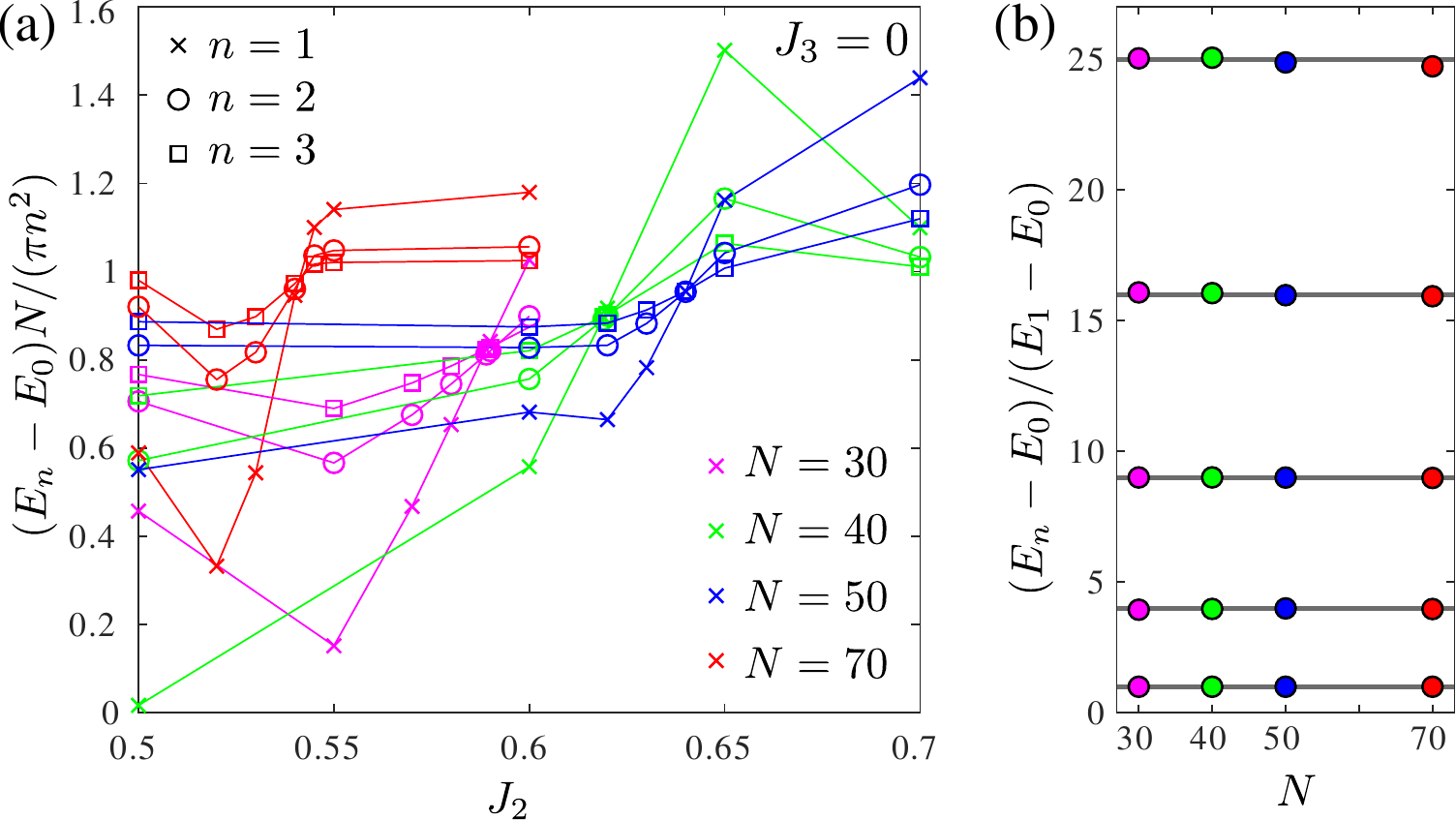}
\caption{(a) Effective velocities across the transition between partially dimerized and floating phases extracted from the gap between various energy levels and the ground state as a function of $J_2$ for fixed value of $J_3 = 0$ and different system sizes. Only first crossing is shown. (b) Conformal towers of states extracted at the crossing points in (a). The DMRG data (circles) agrees with the CFT prediction for WZW SU(2)$_1$ (gray lines) within 2$\%$.}
\label{fig:KT2_velocity}
\end{figure}

In Fig.~\ref{fig:KT2_velocity}(a) we plot the effective velocities extracted from the finite-size spectrum along the line $J_3=0$. As in the previous case, we expect the velocities to cross at the points where the conformal towers are restored. Luckily enough, the logarithmic corrections grow slowly with $J_2$, so for every system size $N$ we observe a very clear crossing of all three lines. 

In order to show that the observed crossings are not a coincidence, but indeed signal the WZW SU(2)$_1$ critical theory, we look at the higher excited states. In Fig.~\ref{fig:KT2_velocity}(b) we compare the structure of the excitation spectrum at the crossing point with the CFT prediction for WZW SU(2)$_1$. Since the position of the crossing changes  with the size of the chain, one cannot expect the effective velocity to remain the same. Therefore, we extract an effective conformal level with respect to the singlet-triplet gap for each system size $N$ as $n_{eff}=(E_n-E_0)/(E_1-E_0)$. This mean that the lowest level in Fig.~\ref{fig:KT2_velocity}(b) that corresponds to $n=1$ is trivial and shown only for completeness. The next two levels with $n=4$ and $n=9$ show how well we identified the location of the crossing in Fig.~\ref{fig:KT2_velocity}(a). Finally, the two highest levels provide the results for $S^z_\mathrm{tot}=4$ and $5$ and show how close the spectrum is to the WZW SU(2)$_1$ conformal tower. All $n_{eff}$ extracted here agrees with the CFT prediction within 2$\%$. This agreement is surprisingly good and suggests either that logarithmic corrections grow inside the floating phase very slow, so that none of the observed crossings are essentially affected, or that there is a process associated with incommensurability that compensates logarithmic corrections at the crossing points.

The position of the crossing point scales with the system size in a non-monotonous way, but oscillates within a wide range of parameters. This makes it impossible with the available numerical method to identify accurately the location of the Kosterlitz-Thouless transition in the thermodynamic limit.

\subsection{Phase transition between the floating and fully dimerized phases}

\begin{figure}[h!]
\centering 
\includegraphics[width=0.49\textwidth]{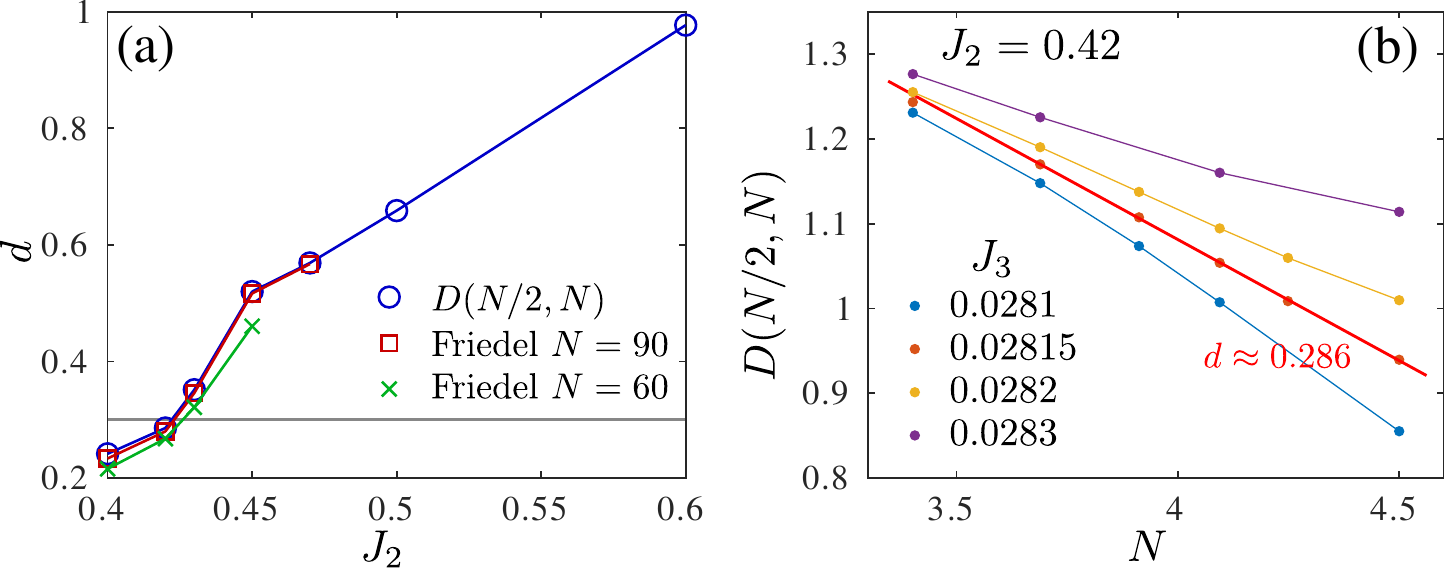}
\caption{(a) Apparent critical exponent along the transition between the floating and fully dimerized phases extracted from a finite-size extrapolation of the middle-chain dimerization (blue circles) and from the fit of the Friedel oscillation profile for two different chain lengths. (b) Example of finite-size scaling of the middle-chain dimerization for $J_2=0.42$ and various values of $J_3$ in the vicinity of the critical point. The slope of the separatrix gives an apparent critical exponent.}
\label{fig:upper_wzw}
\end{figure}

In order to locate the phase transition between the floating and the fully dimerized phases, we look at the finite-size scaling of the dimerization. Despite the presence of algebraic incommensurate order, this method works reasonably well when $J_2$ is not too large. 
An example of such a finite-size scaling is shown in Fig.~\ref{fig:upper_wzw}(b). The slope gives an apparent critical exponent that changes along the transition due to the presence of logarithmic corrections. By analogy with the lower part of the phase diagram we expect the transition between the critical and the fully dimerized phase to be in the WZW SU(2)$_3$ universality class. If this is so, the expected critical exponent (in the absence of log-corrections) take the value $d=0.3$. According to our results shown in Fig.~\ref{fig:upper_wzw} this critical exponent is recovered around $J_2\approx0.42$ and $J_3\approx0.02815$. It is worth mentioning that one might expect significant errorbars in Fig.~\ref{fig:upper_wzw}(a). Although all shown finite-size results are well converged, we cannot reach convergence for chains larger than $N=90$ sites; and it is hard, if actually possible, to estimate finite-size effects due to the presence of quasi-long-range incommensurability. We therefore expect that the location of the end point can be slightly different in the thermodynamic limit, however, the very existence of the end point at which the transition switches from first order to continuous is solid.

\begin{figure}[h!]
\centering 
\includegraphics[width=0.45\textwidth]{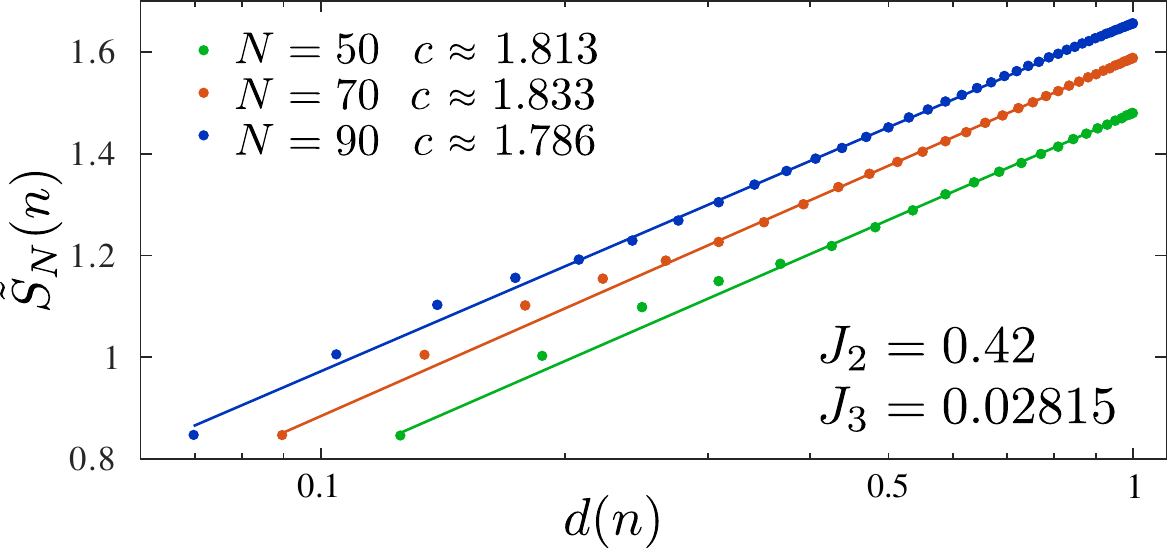}
\caption{Extraction of the central charge for open chains by fitting the reduced entanglement entropy $\tilde{S}_N(n)$ with the Calabrese-Cardy formula of Eq.\ref{eq:calabrese_cardy_obc_corrected_s15} at the end point of the WZW line between the floating and the fully dimerized phases at $J_2\approx 0.42$ and $J_3\approx 0.02815$}
\label{fig:upper_wzw_cc}
\end{figure}

At the end point, where logarithmic corrections are expected to vanish, we extract the central charge from the finite-size scaling of the reduced entanglement entropy as presented in Fig.~\ref{fig:upper_wzw_cc}. The extracted values of the central charge are in very good agreement with the CFT prediction $c=9/5$ for WZW SU(2)$_3$.

\section{Edge states in partially dimerized phase}
\label{sec:edge states}

\subsection{Disappearance of the edge states}

One of the most intriguing and potentially misleading features of the phase diagram is the line where the spin-1/2 edge states disappear. This happens inside the gapped partially dimerized phase and well away from each of the phase boundaries. This is very uncommon; in the original study of $J_1-J_2$ model, the disappearance of the edge states has been interpreted as the indication of a phase transition\cite{roth}. And since the correlation length remains finite at the point where the edge states disappear, it has been suggested\cite{roth} that the transition is first order.

\begin{figure}[h!]
\centering 
\includegraphics[width=0.47\textwidth]{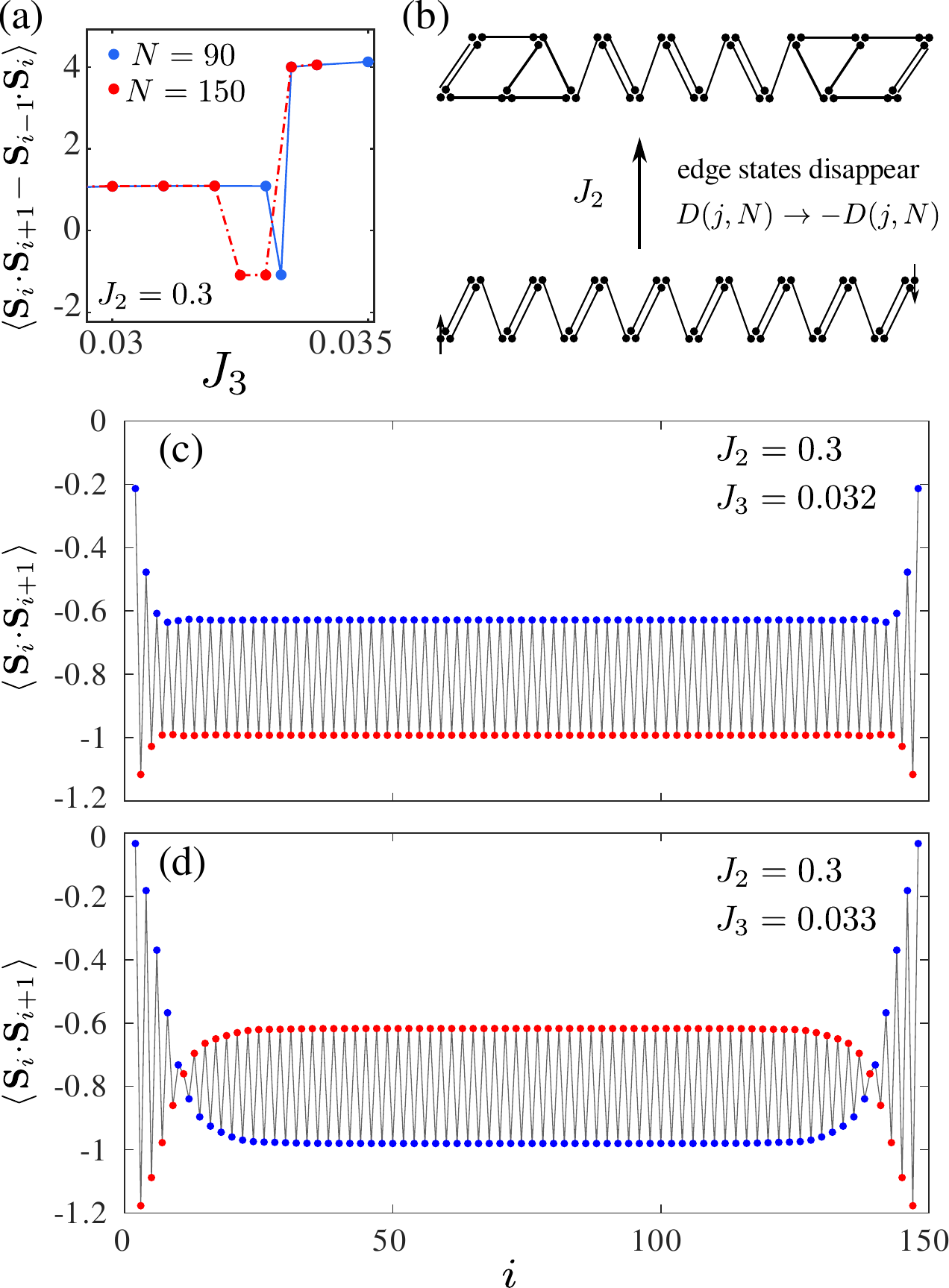}
\caption{(a) Difference between strong and weak consecutive nearest-neighbor correlations computed in the middle of a finite-size chain in the vicinity of the first order transition between the partially and fully dimerized phases. Shortly before the transition, the dimerization switches to a negative value, while its absolute value remains continuous (see Fig.~\ref{fig:dim_ex_s15}(c)). (b) Sketch that show the mechanism of disappearance of the edge states upon increasing the $J_2$ coupling. (c) and (d) Nearest neighbor correlations along a finite-size chain with $N=150$ sites below(c) and above (d) the line where the edge states disappear. Blue (red) dots correspond to odd (even) bonds. One can clearly see an abrupt reorientation of the dimers. }
\label{fig:disap_edge}
\end{figure}

\begin{figure}[h!]
\centering 
\includegraphics[width=0.47\textwidth]{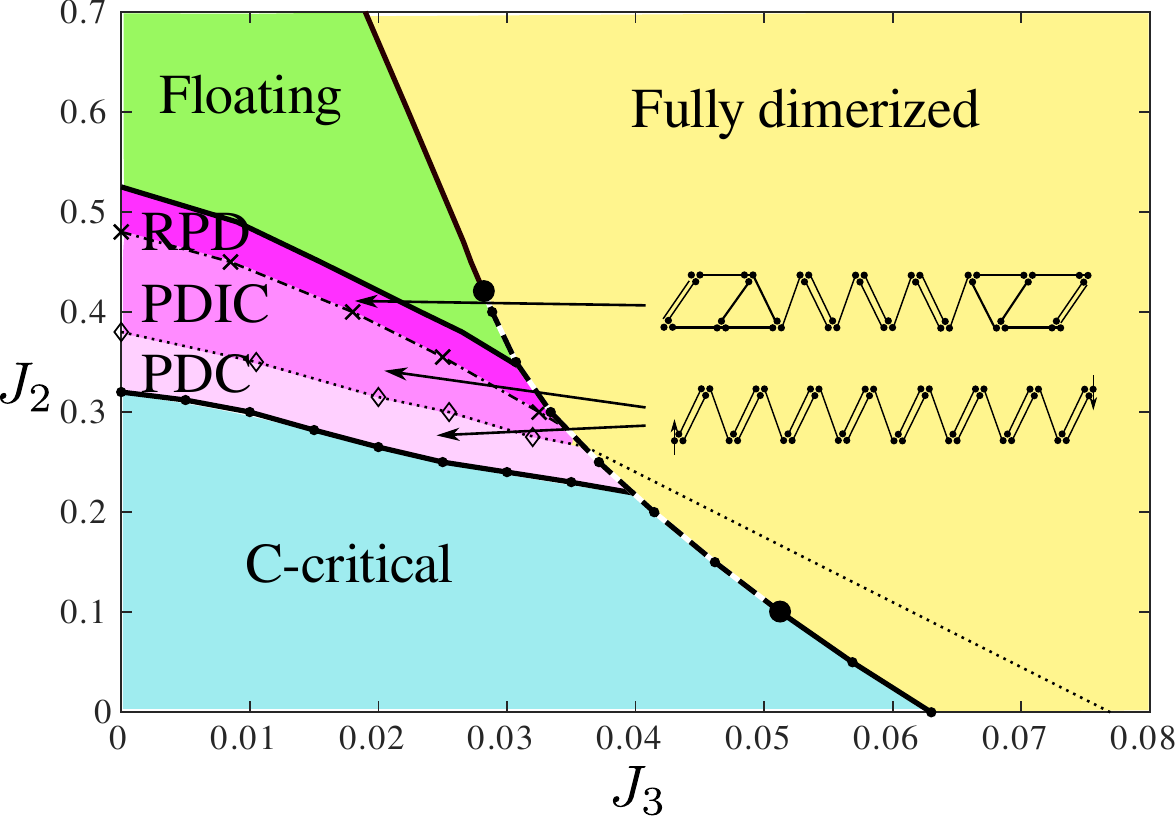}
\caption{Phase diagram with sketches of partially dimerized states with and without edge states.}
\label{fig:pd_edge}
\end{figure}

We would like to propose a different explanation of this phenomenon.
Starting from a certain value of the next-nearest-neighbor coupling, open edges favor domains in a `ladder' state, in which some VBS singlets are located on a few $J_2$ bonds not too far from the edges, as shown in the sketch in Fig.~\ref{fig:disap_edge}(b). It is easy to see that such edge domains can be connected to a domain in the partially dimerized state via non-magnetic domain walls, so that the localized spin-1/2 disappear at both edges. However, this requires a reorientation of dimers: if in the region where the edge states are present, strong dimers are located on every odd bond, then, in the absence of edge states, even bonds becomes stronger. This is well illustrated by Fig.~\ref{fig:disap_edge}(c) and Fig.~\ref{fig:disap_edge}(d). The corresponding regions with and without edge state are marked in the phase diagram of Fig.~\ref{fig:pd_edge}.

Note also that if one keeps track of the sign of the dimerization $D_s(j,N)=\langle \vec S_j\cdot\vec S_{j+1} \rangle - \langle \vec S_{j-1}\cdot\vec S_j \rangle$, there is a small region before the first order transition where the sign of the dimerization changes.

\begin{figure}[h!]
\centering 
\includegraphics[width=0.27\textwidth]{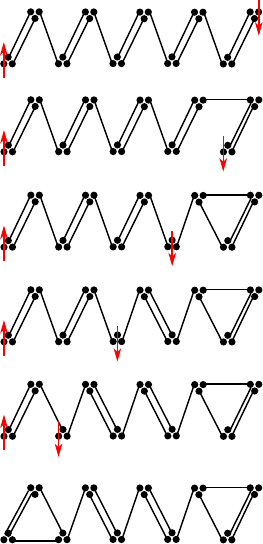}
\caption{Deconfinment of the edge states in the partially dimerized states}
\label{fig:Deconfinment}
\end{figure}

We find it instructive to illustrate the deconfinement of the edge states in the partially dimerized phase using the VBS sketches shown in Fig.~\ref{fig:Deconfinment}. Note that the VBS picture changes only in the vicinity of the edges when the spin-1/2 is moved along the chain. In the bulk, it can move at no energy cost as a domain wall between partially dimerized state with different dimer orientations.  

\subsection{Exact zero modes}

Recently it has been shown that the effective coupling between the spin-1/2 edge states of a spin-1 chain of finite length can be continuously tuned by frustration if it also induces short-range incommensurability\cite{PhysRevB.96.060409}. It implies the existence of several level crossings between the singlet and triplet in-gap  states, i.e. points where the edge states are completely decoupled from each other. Later,this conclusion has been generalized to various spin ladders and to a spin-2 chain\cite{PhysRevB.97.174434}. In all these cases, however, the translation invariance has been preserved.
 
\begin{figure}[h!]
\centering 
\includegraphics[width=0.47\textwidth]{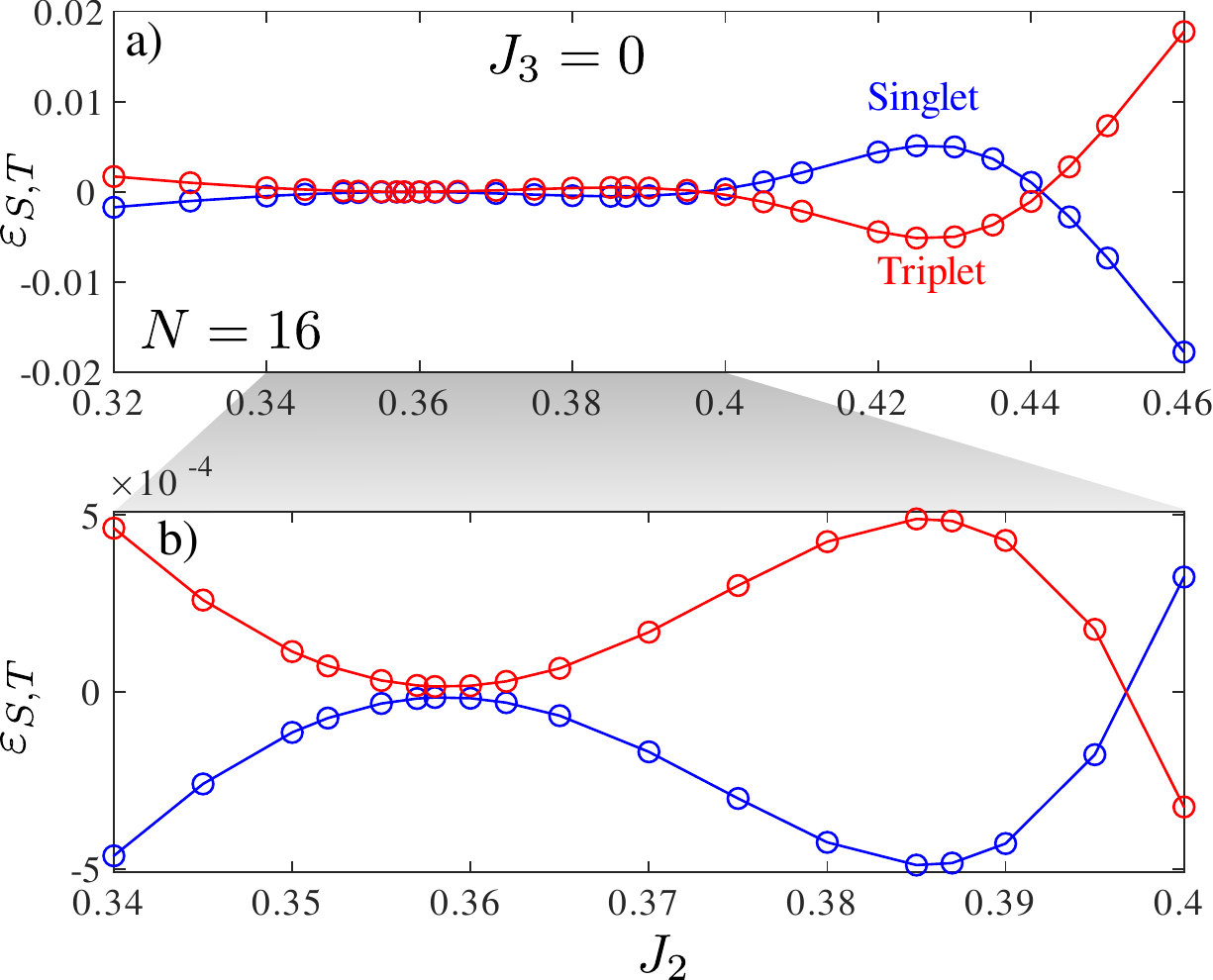}
\caption{Two crossings between singlet and triplet low-lying
energy levels for $N=16$ as a function of the next-nearest-neighbor
coupling constant. Panel (b)  are enlarged parts of (a)}
\label{fig:zeromodes_16}
\end{figure}

\begin{figure}[h!]
\centering 
\includegraphics[width=0.47\textwidth]{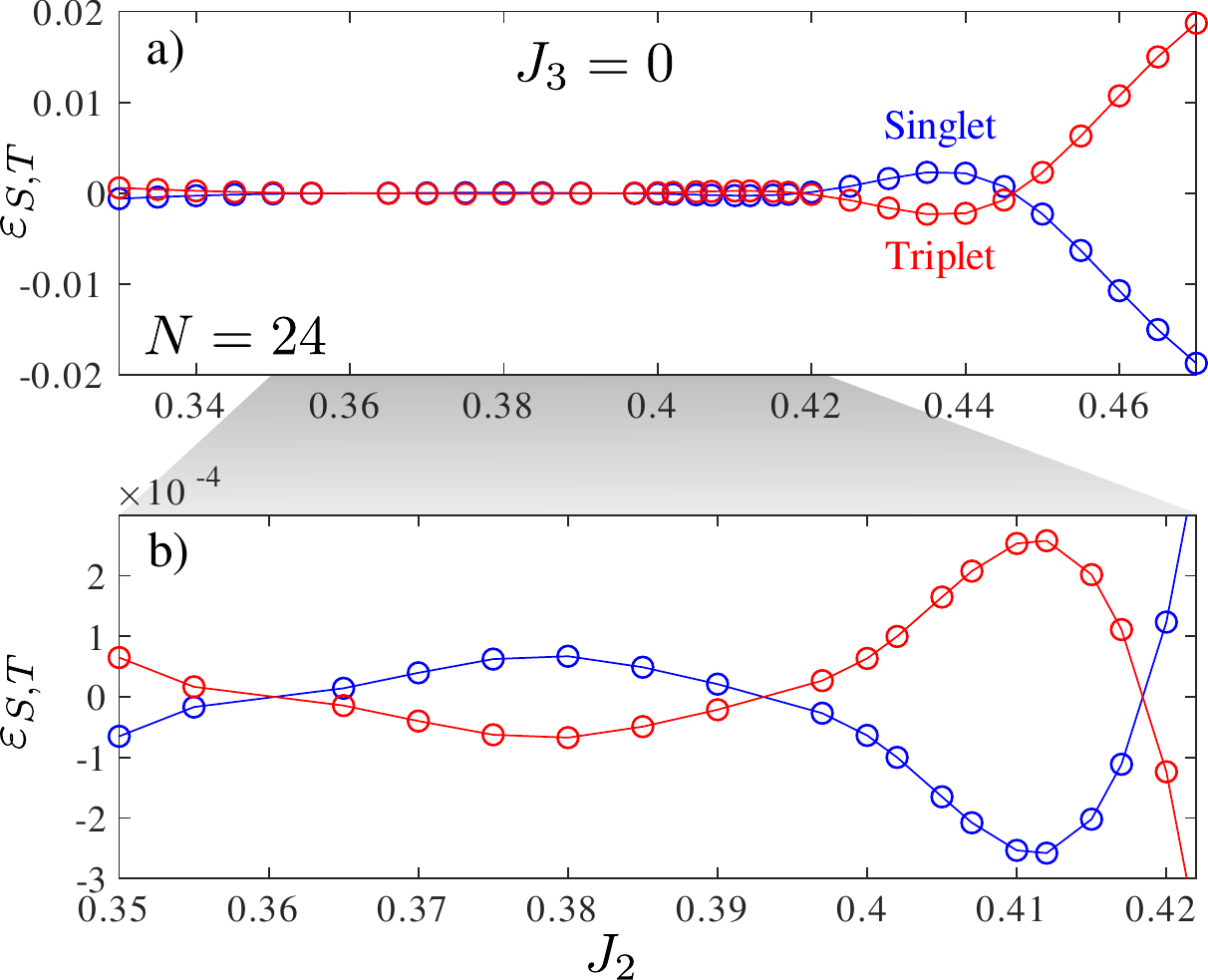}
\caption{Four crossings between singlet and triplet low-lying
energy levels for $N=24$ as a function of the next-nearest-neighbor
coupling constant. Panel (b)  are enlarged parts of (a).}
\label{fig:zeromodes_24}
\end{figure}

In the present case, localized spin-1/2 edges states emerge in the partially dimerized phase, inside which the translation symmetry is spontaneously broken. However, it turns out that the  broken symmetry does not prevent the appearance of exact zero modes. In the region labeled PDIC and located between the disorder line and a line along which the edge states vanish, we observe several level crossings between singlet and triplet in-gap states as shown in  Fig.~\ref{fig:zeromodes_16} and Fig.~\ref{fig:zeromodes_24}.

We have used the average energy as a reference to plot the relative energy of singlet and triplet states: $\varepsilon_{S,T}=E_{S,T}-(E_S+E_T)/2$. So, the level with negative relative energy corresponds to the ground state.
Note that below the disorder line the ground-state is singlet if $N$ is even and it is a triplet if $N$ is odd, while above the line, at which the edge states disappear, the ground-state is always a singlet. It implies that the number of crossings are even for $N$ even and are odd when $N$ is odd.

These results have been obtained  by targeting two states in the sector of $S^z_\mathrm{tot}=0$ as explained in Ref.~\onlinecite{chepiga_dmrg}. Several data points for triplets have been cross-checked by computing the energy of the lowest-energy state in the sector $S^z_\mathrm{tot}=1$.

\section{Conclusions}
\label{sec:conclusion}

To summarize, the  $J_1-J_2-J_3$ model leads to a very rich phase diagram that contains two dimerized phases and two critical phases, to be compared with three phases in the spin-1 case and only two phases in the spin-1/2 case. The combination of both $J_2$ and $J_3$ terms was instrumental to understand the differences between these phases and the transition between them. For instance, the presence of an exactly dimerized line for $J_3$ large enough is very important to support the presence of a first-order transition between the fully dimerized phase and a partially dimerized phase. Still, let us emphasize that all these phases appear in the simple $J_1-J_2$ model, a realistic model that is naturally realized in zig-zag chains. 

This phase diagram reveals a number of unexpected features.
The change in the nature of the phase transition between the c-critical and fully dimerized phase from continuous to first order agrees with our previous results on spin-1 and confirms our prediction that the realized scenario is generic for theories with a marginal operator\cite{j1j2j3_short}. It is nevertheless remarkable because the first order transition appears at the boundary of a critical gapless phase.

Surprisingly enough the upper part of the phase diagram contains a reflected version of this transition - the first order line turns into a continuous WZW SU(2)$_3$ critical line upon increasing the next-nearest-neighbor interaction. We expect only one marginal operator in the theory, so the double change of nature of the critical line suggests that the coupling constant of this marginal operator has a minimum as a function of $J_2$ and therefore crosses zero at the end points. 

The appearance of a floating phase over such a wide parameter range is also quite uncommon. 
Note that an accurate determination of the phase boundaries of the floating phase would require further advances in numerical algorithms. 
 
Finally, we have clarified the origin of the behavior of the edge states in the partially dimerized phase. In particular, we have shown that the disappearance of the edge states does not necessarily imply a phase transition, but can signal local changes of the edges that do not affect the bulk.

Altogether the physics of the frustrated spin-3/2 chain turns out not to be a simple extension of that of the spin-1/2 chain, even if in the absence of frustration they are gapless and described by the same field theory. We hope that the present results will stimulate further experimental investigation in this direction.

\section{Acknowledgments}

This work has been supported by the Swiss National Science Foundation (NC and FM) and by NSERC Discovery Grant 04033-2016 (IA). The calculations have been performed using the facilities of the Scientific IT and Application Support Center of EPFL and computing facilities at the University of Amsterdam.

\bibliographystyle{apsrev4-1}
\bibliography{bibliography}

\end{document}